\documentclass{emulateapj}
\usepackage{apjfonts}
\newcommand{\fluxden}{\mbox{ erg cm$^{-2}$  s$^{-1}$ Hz$^{-1}$ sr$^{-1}$}} 
\newcommand{\hunits}{\mbox{ km s$^{-1}$ Mpc$^{-1}$}} 
\newcommand{\Mpc}{\mbox{ Mpc}}
\newcommand{\Mhz}{\mbox{ MHz}}
\newcommand{\secinv}{\mbox{ s$^{-1}$}}
\newcommand{\kel}{\mbox{ K}}
\newcommand{\mkel}{\mbox{ mK}}
\newcommand{\beq}{\begin{equation}}
\newcommand{\eeq}{\end{equation}}
\newcommand{\beqa}{\begin{eqnarray}}
\newcommand{\eeqa}{\end{eqnarray}}

\newcommand{\tz}{T_0}
\newcommand{\n}{{\hat {\bf n}}}
\newcommand{\p}{{\bf p}}

\newcommand{\hp}{{\hat \psi}}

\newcommand{\dD}{{\delta^D}}
\newcommand{\pp}{{P_\psi}}
\newcommand{\xh}{{x_H}}
\newcommand{\bxh}{{{\bar x}_H}}
\newcommand{\del}{{\Delta^2}}

\def\x{{\bf x}}
\def\k{{\bf k}}
\def\c{{\cal C}}
\def\u{{\bf u}}
\def\r{{ \gamma}}

\begin{document}

\title{21 Centimeter Fluctuations from Cosmic Gas at High Redshifts}

\author{Matias Zaldarriaga\altaffilmark{1}, Steven R. 
Furlanetto\altaffilmark{2}, \& Lars Hernquist\altaffilmark{3}}

\altaffiltext{1} {Harvard-Smithsonian Center for Astrophysics, 60
Garden St., Cambridge, MA 02138; \\ mzaldarriaga@cfa.harvard.edu }

\altaffiltext{2} {Division of Physics, Mathematics, \& Astronomy;
  California Institute of Technology; Mail Code 130-33; Pasadena, CA
  91125; sfurlane@tapir.caltech.edu}

\altaffiltext{3} {Harvard-Smithsonian Center for Astrophysics, 60
Garden St., Cambridge, MA 02138; \\ lars@cfa.harvard.edu }

\begin{abstract}
  
The relatively large Thomson optical depth, $\tau_{es}$, inferred
recently from the WMAP observations suggests that the Universe was
reionized in a more complex manner than previously believed.  However,
the value of $\tau_{es}$ provides only an integral constraint on the
history of reionization and, by itself, cannot be used to determine
the nature of the sources responsible for this transition.  Here, we
show that the evolution of the ionization state of the intergalactic
medium at high redshifts can be measured statistically using
fluctuations in 21 centimeter radiation from neutral hydrogen.  By
analogy with the mathematical description of anisotropies in the
cosmic microwave background, we develop a formalism to quantify the
variations in 21 cm emission as a function of both frequency and
angular scale.  Prior to and following reionization, fluctuations in
the 21 cm signal are mediated by density perturbations in the
distribution of matter.  Between these epochs, pockets of gas
surrounding luminous objects become ionized, producing large HII
regions.  These ``bubbles'' of ionized material imprint features into
the 21 cm power spectrum that make it possible to distinguish them
from fluctuations produced by the density perturbations.
The variation of the power spectrum with frequency can be used to
infer the evolution of this process.  As has been emphasized
previously by others, the absolute 21 cm signal from neutral gas at
high redshifts is difficult to detect owing to contamination by foreground
sources.  However, we argue that this source of noise can be
suppressed by comparing maps closely spaced in frequency,
i.e. redshift, so that 21 cm fluctuations from the IGM can be measured
against a much brighter, but smoothly varying (in frequency)
background.

\end{abstract}


\section{Introduction}
\label{intro}

One of the long-standing goals of cosmology is to understand how
structures have grown through time.  In the usual paradigm, weak
density perturbations were imprinted on the Universe during the
inflationary era.  These grew through gravitational instability,
eventually forming bound halos as well as the cosmic web of sheets and
filaments.  Precise measurements of the cosmic microwave background
(CMB) anisotropies have fixed the initial conditions of the picture
(e.g., \citealt{spergel03}).  The challenge now is to take structure
formation beyond the well-understood linear regime in order to
understand how baryons collapsed into the bound objects that we
observe today, such as galaxies and galaxy clusters, and to understand
how these objects affect their surroundings.  At low or moderate
redshifts ($z \la 6$), galaxies and quasars can be studied in detail
with existing technology.  Unfortunately, the first generations of
protogalaxies are not yet accessible observationally.  Their
properties are nevertheless crucial to understanding both later
generations of galaxies, which form out of these early protogalaxies
in any hierarchical picture of structure formation, and the gross
evolution of baryons in the Universe, because these objects exhibit
strong feedback on their surroundings.  Perhaps the most important
such channel is the reionization of the intergalactic medium (IGM).
When the first protogalaxies or quasars form, they ionize pockets of
surrounding gas.  These \ion{H}{2} regions grow with time and
eventually overlap.  The timing, morphology, and duration of this
event contain a wealth of information about both the ionizing sources
and the IGM (e.g.,
\citealt{wyithe03,cen03,haiman03,mackey03,yoshida03-semian,ybh}).

A great deal of effort has gone into constraining the transition from
a neutral to ionized IGM.  Unfortunately, existing observational
techniques are not optimized to the needed measurements; they have
provided tantalizing constraints on reionization but cannot be used to
map the event in detail.  The most straightforward method is to extend
the ``Ly$\alpha$ forest'' to high redshifts: regions with relatively
large \ion{H}{1} densities appear as absorption troughs in quasar
spectra, which presumably deepen and come to dominate the spectra as
we approach the reionization epoch.  Indeed, spectra of $z \sim 6$
quasars selected from the Sloan Digital Sky Survey\footnote{See
http://www.sdss.org/.}  (SDSS) show at least one extended
region of zero transmission \citep{becker}, indicating that the
ionizing background is rising at this time \citep{fan}.
However, the optical depth of the IGM to Ly$\alpha$ absorption is
$\tau_{Ly\alpha} \approx 6.45 \times 10^5 \xh [(1+z)/10]^{3/2}$
\citep{gp}, where $\xh$ is the neutral fraction.  A neutral
fraction $\xh \ga 10^{-3}$ will therefore render the absorption
trough completely black; quasar absorption spectra can clearly probe
only the latest stages of reionization.  

A second constraint comes from the effects of the ionized gas on the
CMB.  The free electrons Thomson scatter the CMB photons, washing out
the intrinsic anisotropies but generating a polarization signal.  The
total scattering optical depth $\tau_{\rm es}$ is proportional to the
column density of ionized hydrogen, so it provides an integral
constraint on the reionization history.  Recently, the \emph{Wilkinson
Microwave Anisotropy Probe}\footnote{See http://map.gsfc.nasa.gov/.}
(\emph{WMAP}) used the polarization signal to measure a large
$\tau_{\rm es}$, indicating that reionization began at $z_r \ga 14$
\citep{kogut03,spergel03}.  More detailed information on the
reionization history could be obtained by measuring the (large)
angular scales over which CMB polarization is generated
\citep{zal97,kaplinghat03,hu03} or the (small) scales over which
secondary anisotropies are generated by the patchiness of reionization
\citep{gruzinov98,knox98}, but these signals promise to be difficult
to extract \citep{holder03,santos03}.  A third constraint comes from
measurements of the temperature of the Ly$\alpha$ forest at $z \sim
2$--$4$, which suggest an order unity change in the ionized fraction
at $z_r \la 10$ \citep{theuns02-reion,hui03}, although this argument
depends on the timing and history of \ion{He}{2} reionization (e.g.,
\citealt{sokasian02}).

Taken together, these three sets of observations imply a complex
reionization history extending over a large redshift interval ($\Delta
z \sim 10$).  This is inconsistent with a ``generic'' picture of fast
reionization (e.g., \citealt{barkana01}, and references therein).  The
results seem to indicate strong evolution in the sources responsible
for reionization, and a detailed measurement of the reionization history
would contain a rich set of information about early structure
formation \citep{sokasian03a,wyithe03,cen03,haiman03}.
The optimal reionization experiment would: (1) be sensitive to order
unity changes in $\xh$ (to probe the crucial middle stages of
reionization), (2) provide measurements that are well-localized along
the line of sight (rather than a single integral constraint), and (3) not
require the presence of bright background sources, which may be rare
at high redshifts.  The most promising candidate proposed to date is
the 21 cm hyperfine transition of neutral hydrogen in the IGM
\citep{field58,field59a}, which fulfills all three of these criteria.
So long as the excitation temperature $T_S$ of the 21 cm transition in
a region of the IGM differs from the CMB temperature, that region will
appear in either emission (if $T_S > T_{\rm CMB}$) or absorption (if
$T_S < T_{\rm CMB}$) when viewed against the CMB.  Variations in the
density of neutral hydrogen (due either to large-scale structure or to
\ion{H}{2} regions) would appear as fluctuations in the sky brightness
of this transition.  Because it is a line transition, the fluctuations
can also be well-localized in redshift space.  Thus, in
principle, high resolution observations of the 21 cm transition in
both frequency and angle can provide a three-dimensional map of
reionization.  Together with radio absorption spectra of bright
background sources (which can probe much smaller physical scales in
the IGM; \citealt{carilli,furl-21cm}), these observations promise to
shed light both on the early growth of structure and on reionization.

\begin{figure*}
\plotone{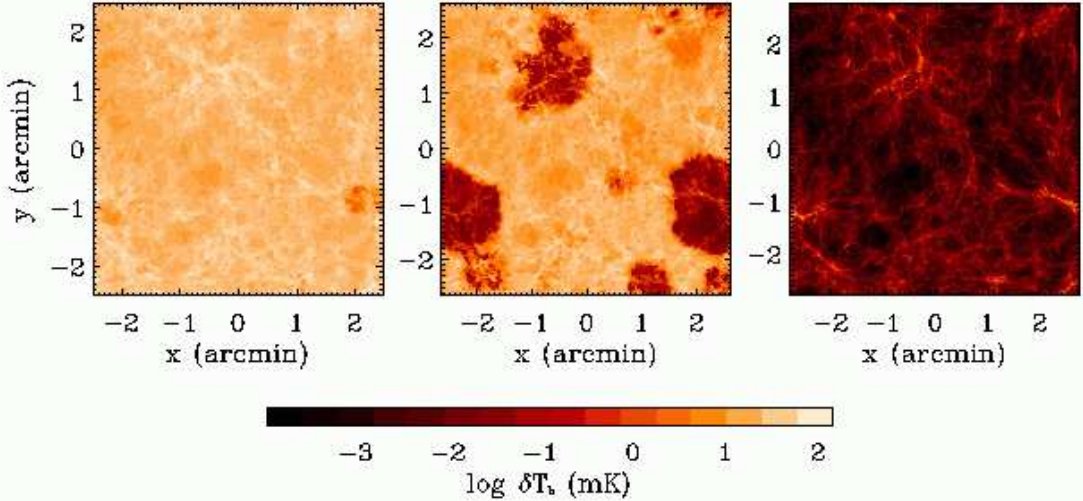}
\caption{ The brightness temperature of the 21 cm transition at
several redshifts, as predicted by the ``late reionization''
simulation analyzed in \citet{furl-21cmsim}.  Each panel corresponds
to the same slice of the simulation box (with width $10 h^{-1}$ comoving Mpc
and depth $\Delta \nu=0.1 \Mhz$), at $z=12.1$, 9.2, and 7.6, from left
to right. The three epochs shown correspond to the early, middle, and
late stages of reionization in this simulation.  (For details about
the simulations, see \citealt{sok01,sh03a,sh03b}.)}
\label{fig:sim}
\end{figure*}

The physics of this transition has been well-studied in the
cosmological context.  Early work focused on fluctuations due to
large-scale structure \citep{scott,kumar,mmr,tozzi,iliev}, because the
signals could be estimated through linear cosmological perturbation
theory.  \citet{shaver} were the first to explicitly consider the
signal at reionization, although they focused on the ``all-sky''
signature rather than the fluctuations.  Recently, \citet{ciardi03}
and \citet{furl-21cmsim} used numerical simulations of reionization to
estimate how the fluctuations would behave during that epoch.  We show
in Figure \ref{fig:sim} three time slices from the simulation analysis
described by \citet{furl-21cmsim}, corresponding to the early, middle,
and late stages of reionization (from left to right).  It is clear that 
both the mean signal and the fluctuations drop abruptly.  Interestingly,
the fluctuations during reionization have a very different morphology
than those due to large-scale structure; the spectrum of fluctuations
thus has the potential to constrain the process of reionization.

In this paper, we present a new approach to 21 cm fluctuations.  We
draw an analogy between these measurements and those of the CMB: in
both cases we wish to measure the level of inhomogeneity as a function
of scale on the sky.  Previous treatments of the 21 cm signal have
focused on measuring fluctuations on a particular patch of the sky,
implicitly referring to imaging observations.  Here we show that a
statistical treatment of the fluctuation power spectrum contains a
great deal of information about reionization.  Furthermore, a large
set of tools for CMB predictions and data analysis has already been
developed (see \citealt{hu02} for a review), so there is much to be
gained by connecting the two.  Indeed, some steps in this direction
have already been taken by \citet{pen03} and \citet{cooray03}, both of
whom considered the effects of lensing on the 21 cm signal.  The
analogy with the CMB is not perfect, however, because the 21 cm signal
can be separated in redshift space; in other words, we can make maps
at a series of frequencies, each of which samples an independent
volume.  In this sense the analogy is closer to redshift surveys
(e.g., \citealt{peebles80}) or weak lensing tomography (e.g.,
\citealt{hu99}).  We therefore develop our formalism with explicit
consideration of how multifrequency information can be used with power
spectrum statistics, in effect generalizing the methodology used to
analyze CMB anisotropies.  After reviewing the physics of the 21 cm
transition in \S 2, we show how to compute the angular power spectrum
of 21 cm fluctuations in \S 3.

In \S 4 we show some simple applications of our approach.  We give
predictions for the angular power spectrum of 21 cm fluctuations from
a fully neutral medium and for a simple toy model of reionization.  In
the former case, fluctuations in the signal are due only to
large-scale structure.  We show that in this regime 21 cm measurements
essentially yield the power spectrum of density fluctuations (see also
\citealt{pen03}).  We then show that variations in the neutral
fraction during reionization distort the power spectrum.  

Another advantage of our approach is that the angular power spectra
are closely related to the physically observed quantities.  This is
especially true for the interferometers that will most likely be used
to measure the redshifted 21 cm signal.  Our results thus
connect theoretical predictions to the potential observations.  For
example, inhomogeneities in the 21 cm signal must be separated from
fluctuations in any foreground sources.  This is particularly
important because the absolute foreground signal will swamp the 21 cm
signal by many orders of magnitude.  While Galactic foregrounds are
expected to be fairly smooth on the relevant angular scales, faint
radio galaxies, starbursts, and even the galaxies responsible for
reionization fluctuate strongly on arcminute scales and dominate those
of the 21 cm signal by at least an order of magnitude
\citep{dimatteo,oh03}.  These results have been used to argue that the
prospects for large angular-scale measurements of the reionization
epoch are dim.  However, both \citet{dimatteo} and \citet{oh03} also
pointed out that all of the (known) foreground sources have
featureless power-law spectra.  Both suggested that the foregrounds
could therefore be removed in frequency space.  As an example,
consider the simple case in which every foreground source has the
\emph{same} spectral index.  Then the foregrounds between two maps at
nearby frequencies would be exactly correlated, while the 21 cm
fluctuations will be uncorrelated because each frequency samples an
independent volume of the IGM.  Comparing two maps closely spaced in
redshift therefore allows one to remove the foreground component.  We
show in \S 5 how the foreground sources can be modeled with our
multi-frequency formalism.  In \S 6 we quantify how well their
contamination can be removed.  We find that foregrounds are much less
important than previously assumed so long as the range of allowed
spectral indices for faint sources is similar to that already measured
for brighter sources.  Finally, we estimate in \S 7 how well the power
spectrum can be measured with the next generation of low-frequency
radio telescopes, and we conclude in \S 8.

When necessary, we assume a $\Lambda$CDM cosmology with
$\Omega_m=0.3$, $\Omega_\Lambda=0.7$, $\Omega_b=0.04$, $H_0=100h
\hunits$ (with $h=0.7$), and a scale-invariant primordial power
spectrum with $n=1$ normalized to $\sigma_8=0.85$ at the present day.

\section{21 cm Radiation from the Intergalactic Medium}
\label{21cm}

The optical depth of a patch of the IGM in the hyperfine transition is
\citep{field59a}
\beqa
\tau & = & \frac{ 3 c^3 \hbar A_{10} \, n_{\rm HI}}{16 
k \nu_0^2 \, T_S \, H(z) } 
\label{eq:tauigm} \\
\, & \approx & 8.6 \times 10^{-3} (1+\delta) \xh \left[
\frac{T_{\rm CMB}(z)}{T_S} \right] 
\left( \frac{\Omega_b h^2}{0.02} \right) \nonumber \\
& & \times \left[ \left(\frac{0.15}{\Omega_m h^2} \right) \, \left(
\frac{1+z}{10} \right) \right]^{1/2}. 
\nonumber
\eeqa
Here $\nu_0=1420.4 \Mhz$ is the rest-frame hyperfine transition
frequency, $A_{10} = 2.85 \times 10^{-15} \secinv$ is the spontaneous
emission coefficient for the transition, $T_S$ is the spin temperature
of the IGM (i.e., the excitation temperature of the hyperfine
transition), $T_{\rm CMB} = 2.73 (1+z) \kel$ is the CMB temperature at
redshift $z$, and $n_{\rm HI}$ is the local neutral hydrogen density.
In the second equality, we have assumed sufficiently high redshifts
such that $H(z) \approx H_0 \Omega_m^{1/2} (1+z)^{3/2}$ (which is
well-satisfied in the era we study, $z > 6$).  The local
baryon overdensity is $1+\delta=\bar{\rho}/\rho$ and $\xh$ is
the neutral fraction.  The radiative transfer equation in the
Rayleigh-Jeans limit then tells us that the brightness temperature of
a patch of the sky (in its rest frame) is $T_b=T_{\rm
CMB}e^{-\tau}+T_S(1-e^{-\tau})$.  We define $\delta T(\nu)$ to be the
observed brightness temperature increment between this patch, at an
observed frequency $\nu$ corresponding to a redshift $1+z=\nu_0/\nu$,
and the CMB:
\beqa
\delta T(\nu) & \approx & \frac{T_S - T_{\rm CMB}}{1+z} \, \tau 
\label{eq:dtb} \\
\, & \approx \, & 23 \, (1+\delta) \xh \left( \frac{T_S - T_{\rm
CMB}}{T_S} \right) \left( \frac{\Omega_b h^2}{0.02} \right) \nonumber \\
& & \times \left[ \left(\frac{0.15}{\Omega_m h^2} \right) \, \left(
\frac{1+z}{10} \right) \right]^{1/2} \mkel.
\nonumber
\eeqa

Assuming that the radiation background includes only the CMB, the \ion{H}{1}
spin temperature is \citep{field58}
\beq
T_S = \frac{T_{\rm CMB} + y_c T_K + y_{{\rm Ly}\alpha} T_{{\rm
      Ly}\alpha}}{1 + y_c +  y_{{\rm Ly}\alpha}}. 
\label{eq:hItspin}
\eeq 
The second term describes collisional excitation of the hyperfine
transition, which couples $T_S$ to the gas kinetic temperature $T_K$.
The coupling coefficient is
\beq 
y_c = \frac{C_{10}}{A_{10}} \, \frac{T_\star}{T_K},
\label{eq:yc}
\eeq 
where $C_{10}(T_K) \propto n_H$ is the collisional de-excitation rate
of the (higher-energy) triplet hyperfine level \citep{allison} and
$T_\star = 2 \pi \hbar \nu_0/k = 0.068 \kel$.  For $T_K \sim 1000
\kel$, the coupling becomes strong when $1+\delta \ga 5 [(1+z)/20]^2$.
The third term in equation (\ref{eq:hItspin}) describes the
Wouthuysen-Field effect, in which Ly$\alpha$ pumping couples the spin
temperature to the color temperature of the radiation field $T_{{\rm
    Ly}\alpha}$ 
\citep{wout,field58}.  We note that $T_{{\rm Ly}\alpha}=T_K$ so long as the
medium is optically thick to Ly$\alpha$ photons \citep{field59b}.
Essentially, the dipole selection rules allow a transition between the
two hyperfine levels of the ground state mediated by the absorption
and subsequent re-emission of a Ly$\alpha$ photon.  The excitation and
de-excitation rates are then controlled by the color temperature of
the radiation field near the line center, which (for a sufficiently
large number of scatterings) must be in thermodynamic equilibrium with
the gas temperature.  The coupling constant for this process is
\beq 
y_{{\rm Ly}\alpha} =
\frac{P_{10}}{A_{10}} \, \frac{T_\star}{T_{{\rm Ly}\alpha}},
\label{eq:yalpha}
\eeq 
where $P_{10}$ is the indirect de-excitation rate of the triplet
level due to absorption of a Ly$\alpha$ photon followed by decay to the
singlet level.  For a diffuse Ly$\alpha$ background, \citet{mmr}
showed that 
\beq 
P_{10} \approx 1.3 \times 10^{-12} J_{-21} \secinv,
\label{eq:j21}
\eeq 
where $J_{-21}$ is the intensity of the background radiation field at
the Ly$\alpha$ frequency in units of $10^{-21} \fluxden$.  Ly$\alpha$
pumping effectively couples $T_S$ and $T_K$ when $J_{-21} \ga 1$.
\citet{ciardi03} argue that $J_{-21} \gg 1$ even at $z \ga 20$.  If
so, $T_S \sim T_K$ throughout the diffuse IGM, even though the
densities are well below the threshold for collisional coupling.

The spin temperature and optical depth therefore depend on the kinetic
temperature of the IGM.  Once Thomson scattering of CMB photons becomes
inefficient at the thermal decoupling redshift $z_d \sim 140$, the IGM
cools adiabatically until the first objects collapse
\citep{couchman86}.  During this era, $T_K<T_{\rm CMB}$.  The cooling
trend reverses itself as soon as significant structure begins to form,
but the subsequent temperature evolution is both inhomogeneous and
highly uncertain.  
While early estimates suggested that Ly$\alpha$ photons themselves
would inject significant thermal energy into the IGM, \citet{chen03}
showed that this heating channel is in reality quite slow.  Instead,
X-rays (primarily from supernovae or accreting black holes) and shocks
are likely to control the temperature evolution of the IGM.  
We expect shocks to heat overdense structures like
sheets, filaments, and virialized halos to $T_K>T_{\rm CMB}$ 
and radiative feedback from stars and quasars to heat the rest of the
gas.  Most estimates suggest that the two processes will rapidly heat
the IGM to $T_K>T_{\rm CMB}$ \citep{venkatesan01,chen03,gnedin03}.
The topology of the two cases of course differs; shock heating will
tend to exaggerate brightness temperature differences by separating
warm, dense regions from cool voids, while X-ray heating will induce a
smooth temperature distribution.

In developing our formalism, we allow $\delta T(\nu)$ to depend on the
parameters $\delta,\xh,$ and $T_S$.  This is the most general case
possible (once the cosmological parameters are fixed) and allows one
to incorporate the full range of physics when necessary.  However, the
arguments above suggest that the situation most relevant to
observations has $T_S \sim T_K \gg T_{\rm CMB}$.  In this limit, the
temperature factor in equation (\ref{eq:dtb}) approaches unity and the
signal is independent of $T_S$.  Thus we essentially assume an era of
significant X-ray heating; in this scenario, the extra heating in
shocked dense regions can be ignored.  For simplicity, we will
restrict ourselves to this case for the illustrative examples in \S 4.
We emphasize that this is, however, an important assumption.  If, for
example, significant heating does not occur until the early stages of
reionization, $\delta T_b$ will have a much more complicated
distribution than we consider here.

Finally, to orient the reader, we note that an observed bandwidth
$\Delta \nu$ corresponds to a comoving distance
\beq
L \approx 1.7 \left( \frac{\Delta \nu}{0.1 \Mhz} \right) \left(
\frac{1+z}{10} \right)^{1/2} \left( \frac{\Omega_m h^2}{0.15}
\right)^{-1/2} \Mpc,
\label{eq:lcom}
\eeq
while a given angular scale $\Delta \theta$ corresponds to 
\beq
R \approx 1.9 \left( \frac{\Delta \theta}{1'} \right)
\left( \frac{1+z}{10} \right)^{0.2} h^{-1} \Mpc
\label{eq:rcom}
\eeq 
over the relevant redshift range.  

\section{Basic Formalism}

We now show how to compute the angular power spectrum of 21 cm
fluctuations.  Unfortunately, a given patch of the sky observed with frequency
bandwidth $\Delta\nu$ does not correspond directly to a physical
volume of the Universe because the observation is performed in
redshift space:  peculiar velocities can move a parcel of gas into or
out of this channel.  However, redshift space distortions will be
unimportant if $\Delta\nu/\nu > v/c$ where $v$ is the typical random
bulk velocity of the gas.  The importance of redshift space
distortions is determined by the ratio
\beq 
{\cal R} =
{\Delta\nu /\nu \over v/c} \approx 20 \times \left [ {(1+z) \over
11}\right ] \times \left [{\Delta \nu \over 0.2 \ {\rm MHz}} \right ]
\times \left [ { v/c \over 10^{-4}}\right ]^{-1} 
\eeq 
For large values of ${\cal R}$, redshift distortions have only a
marginal effect.  \citet{furl-21cmsim} have shown that, for typical
survey geometries, redshift space distortions amplify the signal by at
most $\sim 25\%$ (see also \citealt{tozzi}).  For simplicity we will
ignore them in what follows.  In future work, we will examine the
significance of redshift distortions in detail using numerical
simulations and asses what cosmological information can be extracted
from their detection.\footnote{In principle, if one knows the dark
matter power spectrum, comparison with the observed spectrum could
allow one to extract the peculiar velocities directly.  This procedure
would be easiest when fluctuations in $x_H$ can be ignored, i.e. in
the very early stages of reionization.}

If redshift distortions can be neglected, there is a one-to-one
correspondence between frequency and redshift. The bandwidth of the
experiment is characterized by some response function $W(\nu)$. The
observed brightness temperature is of the form
\beq\label{tdef}
T(\n,\nu)=\tz(r_0) \int dr W_{r_0}(r)   \psi(\n,r),
\eeq
where $\n$ is the direction of observation, $\tz$ is a normalization
constant which depends on redshift and $\psi(\n,r)$ is the
dimensionless brightness temperature, $\psi(\n,r)=(1+\delta) x_H
(T_S-T_{\rm CMB})/T_S$.  Note that $\delta T(\nu)=T_0(r_0) \psi(\n,r)$
in equation (\ref{eq:dtb}).  The projection window $W_{r0}(r)$ is a
function peaked at $r_0$, the radial distance corresponding to the
observed frequency $\nu$, and has a width $\delta r$.

We can expand $\psi(\x)$ as a Fourier series, 
\beqa\label{psik}
\psi(\x)&=&\int {d^3k \over (2\pi)^3} \ \hp(\k) e^{i\k\cdot \x}
\nonumber \\ &=& \int {d^3k \over (2\pi)^3} \ \hp(\k) \sum_{lm} 4 \pi
i^l j_l(kr) Y_{lm}^*(\hat \k) Y_{lm}(\n),  
\eeqa 
where $j_l(x)$ are the spherical Bessel functions and $Y_{lm}(\n)$ are
the spherial harmonics.  The statistics of $\psi$ are determined by its power 
spectrum,\footnote{Actually, $\psi$ is unlikely to be a true Gaussian
  random field because of the distribution of $x_H$, so its statistics
  are not \emph{entirely} determined by $P_\psi(k)$.}
\beq
\langle \hp(\k_1) \hp(\k_2) \rangle = (2\pi)^3 \dD(\k_1+\k_2)
\pp(k_1).  
\eeq

We use equations (\ref{tdef}) and  (\ref{psik}) to calculate the
spherical harmonic decomposition $a_{lm}$ of the observed temperature: 
\beqa
a_{lm}(\nu)&=&  4 \pi i^l \int {d^3k \over (2\pi)^3} \ \hp(\k)
\alpha_l(k,\nu) Y_{lm}^*(\hat \k), \nonumber \\
\alpha_l(k,\nu)&=&\tz(r_0) \int dr W_{r_0}(r)  j_l(kr) .
\eeqa

We can define the angular power spectrum as 
\beqa\label{cldef}
\langle a_{l_1 m_1}(\nu_1) a^*_{l_2 m_2}(\nu_2)\rangle&=&\delta_{l_1
  l_2}  \delta_{m_1 m_2} C_{l_1}(\nu_1,\nu_2) \nonumber \\ 
C_{l}(\nu_1,\nu_2)&=&4 \pi  \int {d^3k \over (2\pi)^3} \pp(k)
\alpha_l(k,\nu_1) \alpha_l(k,\nu_2) \nonumber \\ 
&=& 4\pi \int {dk \over k } \Delta^2_{\psi}(k) \alpha_l(k,\nu_1)
\alpha_l(k,\nu_2) . 
\eeqa
Here $\Delta^2_{\psi}(k)=k^3 P_{\psi}/(2\pi^2)$ and we have used the
isotropy of $P_{\psi}$.  This formula encodes both the case of the
power spectrum of maps at one particular frequency (when
$\nu_1=\nu_2$) as well as the correlations between maps at different
frequencies.  Equation (\ref{cldef}) forms the basis for the analysis
that makes it possible to relate the 21 cm fluctuations to the
evolution of the ionization state of the IGM.  Note that the $C_l$'s
approach zero as $\nu_1$ and $\nu_2$ depart from each other  
for two reasons:  because the frequency-space window functions no
longer overlap in this limit \emph{and} because the fluctuations in
the IGM are uncorrelated on large scales.
The latter property can be used to separate the 21 cm signal from
contamination by foreground sources that vary smoothly with frequency.

We can investigate the behavior of equation (\ref{cldef}) by
considering various limits. We want to understand how equation
(\ref{cldef}) depends on the width of the response function, $\delta
r$ (which describes the bandwidth of the observation). The figure of
merit is $l \delta r /r$
(i.e. the ratio of the radial to transverse scales probed by the
observation).  
We first consider the limit in which the
response function can be considered to be a delta function, $l \delta
r/r \ll 1$. In that case,
\beq
\alpha_l\approx \tz(r_0) j_l(k r_0),
\eeq
so that 
\beqa
C_{l}(\nu,\nu)&\approx& 4\pi \tz^2(r_0)\int {dk \over k}
\Delta^2_{\psi}(k) j_l^2(kr_0). 
\eeqa
In the limit in which the power spectrum can be approximated by a
power law, $\Delta^2_{\psi}(k)=(k/k_*)^n$ we then have
\beqa
C_{l}(\nu,\nu)&\approx&4\pi \tz^2(r_0)\Delta^2_{\psi}
(l/r_0) \int {dx \over x} (x/l)^n j_l^2(x) \nonumber \\
{l(l+1)C_{l}(\nu,\nu) \over 2 \pi}&\approx&\tz^2(r_0) \Delta^2_{\psi}(l/r_0)
f(n) \nonumber  \\
f(n,l)&=&  {  \sqrt{\pi} (l+1) \Gamma[1-n/2] \Gamma[l+n/2] \over 2
  l^{n-1}  \Gamma[(3-n)/2]\Gamma[2+l-n/2]}. 
\eeqa
With this definition $f(n,l)$ is a very weak function of both $l$ and
$n$ and it is normalized so that $f(0,l)=1$. Thus in this limit
$(l\delta r /r \ll 1)$, the temperature fluctuations simply trace the
underlying $\psi$ fluctuations,  
\beq\label{approx1}
{l(l+1)C_{l}(\nu,\nu) \over 2 \pi}\propto \tz^2(r_0)\Delta^2_{\psi}(l/r_0) .
\eeq

We now consider the opposite regime, $l \delta r/r \gg 1$, the large
bandwidth or Limber limit \citep{limber,peebles80}.  We can
approximate the integral in a different way, 
\beqa
C_{l}(\nu,\nu)&=&4\pi  \int {d^3k \over (2\pi)^3} \pp(k)
\alpha_l^2(k,\nu) \nonumber \\ 
&=&4\pi  \tz^2 \int dr_1 W(r_1)\int dr_2 W(r_2) \nonumber \\
& & \times \int {4 \pi k^2 dk
  \over (2 \pi)^3} \pp(k) j_l(k r_1) j_l(k r_2).
\eeqa
The Bessel functions $j_l(x)$ are very small for $x<l$ and start to
oscillate when $x\sim l$. Thus, the integral over $k$ will receive
contributions only from a region around $k \sim l/r$ with width of
order $\Delta k \sim 1/\delta r$. Modes with $k > k+\Delta k $ will be
out of phase for typical separations of the two points $r_1$ and
$r_2$. In this regime we can approximate $\pp(k)\approx
\pp(l/r_1)$. The integral over $k$ is then proportional to
$\dD(r_1-r_2)/r_1^2$ so that
\beqa\label{approx2}
C_{l}(\nu_1,\nu_2)&=& \tz^2 \int dr W^2(r) {P(l/r) \over r^2} \nonumber \\
{l(l+1)C_{l}(\nu,\nu) \over 2 \pi}&\propto&\tz^2  \Delta^2_{\psi}(l/r_0)
{r_0 \over l \delta r}. 
\eeqa
This is the standard Limber's equation, widely used in the context of
weak lensing (e.g. \citealt{kaiser92}).
 
Equations (\ref{approx1}) and (\ref{approx2}) are easy to understand.
For a sufficiently narrow frequency response ($l\delta r/r \ll 1$),
the angular fluctuations directly trace those of the underlying $\psi$
field. However, for a surface of finite width $\delta r$ only those
modes with radial $k \la 1/\delta r$ can contribute because the
response function averages out larger $k$ modes.  Thus when the
surface becomes too thick in the radial direction, angular
fluctuations are no longer of order $\sim k^3 P(k)$ but become $\sim
k^2 P(k)/\delta r$, as seen in equation (\ref{approx2}).

To estimate the $l$ at which the width of the surface begins to damp
the fluctuations, we can consider an Einstein de-Sitter universe in which
$a(\tau)=(\tau/\tau_0)^2$, where $a$ is the scale factor and $\tau$ is
the conformal time.  In this case $\delta \tau /\tau = (1/2) \delta a
/a=(1/2) \Delta \nu/\nu $. Thus, the changes in radial distance
$r=\tau_0-\tau$ are just
\beq\label{delrr}
|{\delta r \over r}| \approx{1\over 2 \sqrt{1+z}} \  {\Delta \nu  \over \nu},
 \eeq 
where $\nu$ is the observed frequency, $\nu=\nu_0/(1+z)$.  
For $\Delta \nu = 0.2 \  {\rm MHz}$ the corresponding value of $l$ from
equation (\ref{delrr}) is $l\sim 5000$, or arcminute scales. 
The damping of fluctuations for large bandwidths has important
implications for the choice of $(\Delta \nu,l)$ in a given observation
(see Figure 4 below and the discussion thereof).

\section{Simple Model for the Correlations of the Brightness Temperature}

In this section we will make a simple model for the correlations of
the dimensionless brightness temperature, $\psi$. A more detailed
study using simulations is left for future work. 

We assume that $T_S \gg T_{\rm CMB}$ so that $\psi=x_H (1+\delta)$. To
calculate the power spectrum of $\psi$ we need a model for the
correlations of the neutral fraction, $x_H$.  We will follow a similar
treatment as that used to model the effect of patchy reionization
on CMB anisotropies (e.g. \citealt{gruzinov98,knox98}). For
simplicity, we will model the the fluctuations in $x_H$ as if they
were produced by a set of uncorrelated ``bubbles" of typical size
$R$. We denote the average value of $\xh$ as $\bxh$.  We allow both
$R$ and $\bxh$ to depend on redshift.  Under this simplifying
assumption we can model the correlations as
\beq
\langle x_H(\x_1) x_H(\x_2)\rangle = \bxh^2 + (\bxh - \bxh^2) f(x_{12}/R) ,
\eeq
where $f(x)$ is a function with the following limits: $f(x) \approx 1$
for $x \ll 1$ and $f(x) \approx 0$ for $x \gg 1$. The details of this
function depend on the shape of the bubbles. If the bubbles are
correlated, one should interpret $R$ as an effective size that
depends on the correlation length between distinct bubbles. To calculate
observables, we will take $f(x)=\exp(-x^2/2)$.

The correlations of $\psi$, $\mu(x_{12}) \equiv \langle \psi(\x_1)
\psi(\x_2)\rangle  - \langle \psi \rangle^2$, become 
\beqa\label{corr}
\mu(x_{12}) & = & [ \bxh^2+(\bxh - \bxh^2) f(x_{12}/R)] \xi(x_{12}) +
\nonumber \\ 
& & (\bxh -
\bxh^2) f(x_{12}/R)+\eta(x_{12})[2 \bxh + \eta(x_{12})], 
\eeqa
where $\xi(x_{12})=\langle \delta(\x_1) \delta(\x_2)\rangle$ is the
correlation function of the density field and $\eta(x_{12})= \langle
\delta(\x_1) x_H(\x_2)\rangle$ gives the cross correlation between the
density and neutral fraction fields. To keep things as simple as
possible we will ignore this last term in what follows.  We will
explore the consequences of including it in future work.  
The correlation function is related to the density power spectrum by
\beq 
\xi(r)=\int {dk \over k} \del_\rho(k){\sin kr\over kr}.  
\eeq

Equation (\ref{corr}) has the following limits:
\beqa
\mu(x_{12}) &\approx& \bxh  \xi(x_{12}) + (\bxh - \bxh^2) \ \
(x_{12} \ll R), \nonumber \\ 
\mu(x_{12}) &\approx& \bxh^2  \xi(x_{12})  \ \  \ \ \ \  \ \ \ \ \ \ \
\ \ \ \ \ \  (x_{12} \gg R). 
\eeqa

When $x_{12} \gg R$ both points are basically independent as far as
the correlations in $x_H$ are concerned, so the correlations are given
by those of the density field times the probability that each of the
two points falls in a neutral region, $\bxh^2$. On scales smaller than
the ``bubble" size, both points fall either inside or outside a
``bubble" so only one factor of $\bxh$ multiplies $\xi$. On top of the
fluctuations produced by the density there are those created by the
presence of the bubbles, $(\bxh - \bxh^2)$.  To illustrate the
behavior of equation (\ref{corr}) we can take the correlation function
of the density to be a power law, $\xi(x)=(x/x_0)^{-n}$. Moreover,
assume that $x_0 < R$ and $\bxh \sim 0.5$. On scales smaller than
$x_0$, $\mu(x)\approx \bxh (x/x_0)^{-n}$. In the range $x_0 < x< R$,
$\mu(x)\approx \bxh (1-\bxh)$, while for $x \gg R$, $\mu(x)\approx
\bxh^2 (x/x_0)^{-n}$.  Thus there is a feature in the correlation
function on the scale of the bubbles. In this simple model the $\psi$
correlations trace the matter correlations on both large and small
scales (but with different amplitudes). On scales intermediate between
the size of the bubbles and the non-linear scale the correlation
function flattens out.

\begin{figure}
  \plotone{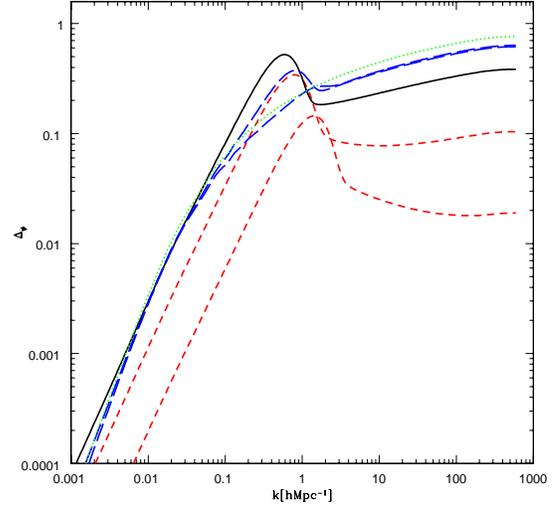}
     \caption{Power spectrum of $\psi$ fluctuations,
     $\Delta_\psi^2=k^3P_\psi/2\pi^2$, in the model described in the
     text. The lines correspond to $z=12,11,10,9$, and $8$. At $z=10$
     (solid black line), $\bxh=1/2$. The long-dashed blue lines
     correspond to $z=12$ and $11$ at the beginning of reionization
     and the short-dashed red lines to $z=9$ and $8$.  The dotted
     green line shows the power spectrum of the density field
     at $z=10$. Note that we use the linear dark matter power spectrum
     in computing the density fluctuations.}
    \label{ppsi}
\end{figure}

We can Fourier transform equation (\ref{corr}) (with $\eta=0$) to
obtain an expression for the power spectrum of $\psi$, 
\beqa\label{powerpsi}
 \del_\psi(k) & = & \bxh^2 \del_\rho(k)+(\bxh - \bxh^2) \del_{x\rho
 }(k) \nonumber \\ & & +(\bxh - \bxh^2)  \del_{x}(k), 
\eeqa
where we have introduced
\beqa
\del_{x}(k)&=&{k^3 {\hat f}(k) \over 2\pi^2}, \nonumber \\
\del_{x\rho}(k)&=&{k^3 \over 2\pi^2} 
\int {d\k^\prime \over (2\pi)^3} P_\rho(\k-\k^\prime) {\hat
  f}(\k^\prime),
\eeqa
with ${\hat f}(k)$ the Fourier transform of $f(x)$ and $P_\rho(\k)$ is
the power spectrum of the density fluctuations.

In Figure \ref{ppsi} we show the power spectrum of $\psi$ calculated
from equation (\ref{powerpsi}).  For illustrative purposes, we used a
simple model for the mean neutral fraction as a function of redshift, 
\beq
\bxh(z)={1 \over 1+ \exp[-(z-z_o)/\Delta z]} ,
\eeq
with $z_0=10$ and $\Delta z=0.5$. For the correlation length $R$ we
used a maximum value of $3 h^{-1}{\rm Mpc}$ when $\bxh=0.5$ and
smaller values when $\bxh$ deviates in both directions from $0.5$ so
as to keep the number density of bubbles fixed.  We show results for,
$(z,\bxh,R)=$ (12,0.98,1.24), (11,0.88,2.24), (10,0.5,3),
(9,0.12,2.24), (8,0.02,1.24). For comparison, we also show the power
spectrum of matter fluctuations at redshift $z=10$.  The form of this
choice for the evolution of the mean neutral fraction is motivated by
numerical simulations of reionization (e.g. Figures 5 and 9 of
\citealt{sokasian03a,sokasian03b}, respectively).  Note that we have
simply used the linear dark matter power spectrum at the appropriate
redshift in computing $\Delta^2_\rho$ for the Figure.  In reality, of
course, we should use the full \emph{gas} power spectrum, which
includes the nonlinear growth of structure on small scale and
smoothing due to the finite pressure of the gas.  The differences are,
however, not large (see, for example, Figure 2 of
\citealt{furl-21cmsim}), so the linear dark matter spectrum will
suffice for our purposes here.  We will consider modifications due to
the true power spectrum in future work.

Before reionization begins, the power spectrum of $\psi$ is simply the
power spectrum of the matter fluctuations. As the neutral fraction
decreases it  develops a feature on the scale of the bubbles, roughly
at $k\sim 2/R$. The feature has maximum amplitude when $\bxh=0.5$. As
the neutral fraction decreases further, the 21 cm fluctuations
disappear because there is no longer any more neutral hydrogen to emit
radiation. Note that when the bubbles appear, the power spectrum on
the largest scales is a power law $\Delta^2_\psi \propto k^3$,
corresponding to Poisson fluctuations. That is, the Poisson
fluctuations induced by the discrete nature of the bubbles dominates
over the fluctuations resulting from $\delta$. 

\begin{figure}
  \plotone{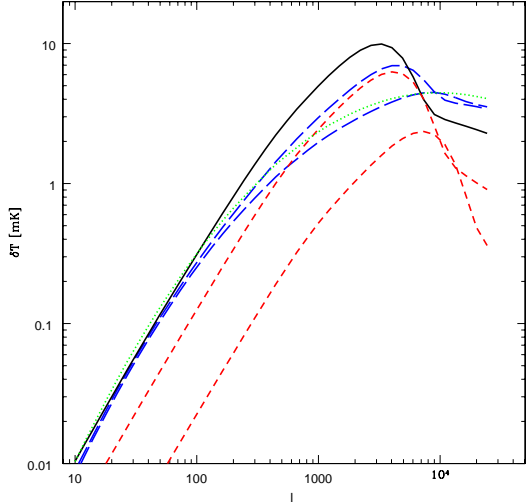}
     \caption{Angular power spectrum of 21cm fluctuations at
     $z=12,11,10,9$, and $8$ in the model described in the text (the
     curves are the same as in Figure \ref{ppsi}).  The bandwidth is
     $0.2 \Mhz$.}
    \label{cl}
\end{figure}

In Figure \ref{cl} we show the corresponding angular power spectrum
$\delta T=[l(l+1)C_l/2\pi]^{1/2}$ with a window function $W_{r_0}(r)$
of Gaussian shape, centered at the appropriate redshift and with a
full width at half maximum of $0.2$ MHz. The angular power spectrum
traces the behavior of the 3D power spectrum shown in Figure
\ref{ppsi}. It develops a feature on the scale of the bubbles, $l\sim
k r\sim 2 r/R$.  We emphasize that the precise shapes of the curves
like those in Figures \ref{ppsi} and \ref{cl} depend on the morphology
of the ionized regions.  Here, we have adopted a simple model in which
these ``bubbles'' are described by spheres whose size evolves simply
with redshift.  In reality, the ionized regions have complicated
morphology and evolution, as indicated by Figure 1, that will likely
imprint more complex features into the power spectra than suggested by
Figures \ref{ppsi} and \ref{cl}.  We will examine these issues further
in due course.

\begin{figure}
  \plotone{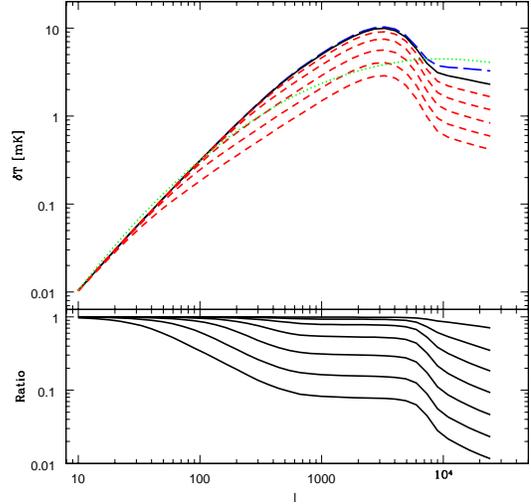}
     \caption{\emph{Top panel:} Angular power spectrum at redshift
$z=10$ for observations done with filters of full width at half
maximum (0.1, 0.2, 0.4, 0.8, 1.6, 3.2, 6.4) MHz.  The dotted line in
the top panel shows the power spectrum of $\delta$ at $z=10$.
\emph{Bottom panel:} Ratio between the power spectra in the top panel
and one corresponding to a delta function filter.}
    \label{clnu}
\end{figure}

On small scales, the angular power spectrum in Figure \ref{cl} changes
slope and stops tracing $\Delta_\psi$. This occurs when the window in
frequency becomes too wide and we enter into the Limber regime
[equation (\ref{approx2})],
where the angular fluctuations are damped by one power of $k$.  To
illustrate this further, we plot the angular power spectrum for
several spectral widths in Figure \ref{clnu}. As the width of the
filter increases, the level of fluctuations decreases.  The Figure
shows that the damping is insignificant until the Limber regime is
reached at $l \delta r/r \sim 1$; in the regime where the fluctuations
are dominated by the bubbles, this happens when $\delta r/R \sim
1$. For bandwidths larger than this, the level of fluctuations scales
as $1/\delta r\propto 1/\Delta \nu$ for a fixed angular scale.  On the
other hand, the signal \emph{per channel} is proportional to the
bandwidth.  Thus, if one is interested in a particular angular scale
$l$, it is best to choose the bandwidth $\Delta \nu$ such that $l
\delta r/r \la 1$.

\section{Modeling the Contaminants}

One of the major challenges in observing this signal is separating it
from the many other (stronger) low-frequency radio sources.  There are
several potential sources of noise, but those associated with foreground
sources are likely to be smooth in frequency space.  In this section
we will show how this smoothness can be used to remove this
contamination.  We will explicitly consider point source foregrounds,
as discussed by \citet{dimatteo} and \citet{oh03}, but our method can
be easily be extended to other types (such as diffuse Galactic
synchrotron emission), provided that their power spectra can be
estimated.

Let us assume that there is a collection of different types of sources
with a range of spectral indices $\zeta$.  We describe them by a
luminosity function $d^2n/dSd\zeta$, which gives the average number of
sources per steradian per unit flux $S$ and spectral index $\zeta$. We
will assume that the sources are clustered but not necessarily that
sources of different spectral indices are perfectly correlated. The
clustering on the sky is described by the angular power spectrum,
\beqa
\langle a^c_{l_1m_1}(\zeta_1) a^c_{l_2m_2}(\zeta_2)\rangle  & = & 
\delta_{l_1 l_2}\delta_{m_1m_2} C_{l_1}^c(\zeta_1,\zeta_2) \\
& = & 
\delta_{l_1 l_2}\delta_{m_1m_2}  \c^c_{l_1}(
\zeta_1,\zeta_2)\sqrt{C_{l_1}^c(\zeta_1)C_{l_1}^c(\zeta_2)} , 
\nonumber
\eeqa
where $C_{l}^c(\zeta_1,\zeta_2)$ is the cross correlation between the
populations and $C_{l}^c(\zeta) =C_{l}^c(\zeta,\zeta)$ is the auto
correlation of a given population. These are nothing more than the
Legendre transforms of the respective correlation functions. We have
also introduced the correlation coefficient $\c^c_{l}(\zeta_1,\zeta_2)
= C_{l}^c(\zeta_1,\zeta_2)/\sqrt{C_{l}^c(\zeta_1)
  C_{l}^c(\zeta_2)}$. 

The power spectrum of fluctuations on the sky produced by these
sources is given by, 
\beqa\label{bigmess} 
C_l(\nu_1,\nu_2)&=&\int dS d\zeta
{d^2n\over dS d\zeta} S^2 \left({\nu_1\over \bar\nu}\right)^\zeta
\left({\nu_2\over \bar\nu}\right)^\zeta \\ &+& C_{l}^c (\bar \zeta)
\int dS_1 d\zeta_1 {d^2n\over dS_1 d\zeta_1} \int \ dS_2 d\zeta_2
{d^2n\over dS_2 d\zeta_2} \nonumber \\ &\times &\c^c_{l}(
\zeta_1,\zeta_2)\sqrt{C_{l_1}^c(\zeta_1) \over C_{l}^c(\bar \zeta)}
\sqrt{C_{l}^c(\zeta_2) \over C_{l}^c(\bar \zeta)} \left({\nu_1\over
\bar\nu}\right)^{\zeta_1} \left({\nu_2\over
\bar\nu}\right)^{\zeta_2}. \nonumber 
\eeqa 
The first term is the Poisson contribution and the second comes from
the clustering of the sources. We have denoted the average spectral
index by $\bar \zeta$.

We will eventually show that what interferes with a measurement of the
21 cm signal is the fact that the foreground maps in different
frequencies may not be perfectly correlated. If the maps were
perfectly correlated, then one could in some sense subtract the map at
one frequency from another and clean the map of all contamination. In
our simple model the presence of sources with several spectral
indices is at the heart of the fact that maps at different
frequencies become uncorrelated.

To make some progress, we adopt some assumptions about the different
quantities that enter into equation (\ref{bigmess}). We will take 
\beq
{d^2n\over dS d\zeta} = {dn\over dS } f(\zeta)=  {dn\over dS } \ \
{e^{-(\zeta-\bar\zeta)^2/2\delta\zeta^2} \over \sqrt{2
    \pi}\delta\zeta} , 
\eeq
where $\delta\zeta$ measures the range of spectral indices of the
sources.  This Gaussian form is a reasonable description of the
spectral index distribution of low-frequency radio sources measured by
\citet{cohen03}.  We need to model the correlation coefficient between
different sources. We know that $\c_l(\zeta,\zeta)=1$ and should decay
as the two $\zeta$'s depart from each other. We will take
\beq
\c(\zeta_1,\zeta_2)=e^{-(\zeta_1-\zeta_2)^2/2\sigma_\zeta^2},
\eeq
where $\sigma_\zeta$ measures how sources with different spectral
indices become uncorrelated.  For simplicity, we will assume that $\c$
is independent of $l$. This could be relaxed easily but would make our
expression a bit more complicated.   To the extent that all sources
are tracing the same underlying distribution of matter, they should be
perfectly correlated, even if their bias is somewhat different. Only
the stochastic part of the bias contributes to the loss of
correlation. Thus we expect $\sigma_\zeta$ to be large compared to
$\delta \zeta$; that is, all the sources should be well correlated on
the sky.  Finally, we will approximate
\beq
\sqrt{C_{l}^s(\zeta) \over C_{l}^s(\bar \zeta)} \approx 1+ {1\over
  2} {d\ln C_l \over d\ln \zeta} ({\zeta\over \bar \zeta} -1), 
\eeq
meaning that we will compute quantities of interest only to lowest
order in the change of clustering with population.  

We use the above formulae to calculate the  correlation coefficient
between maps at different frequencies, 
\beq
I_l(\nu_1,\nu_2)= {C_l(\nu_1,\nu_2) \over \sqrt{C_l(\nu_1,\nu_1)
    C_l(\nu_2,\nu_2)}}.
\eeq
We calculate $I_l(\nu_1,\nu_2)$ as a series in $\ln(\nu_1/\nu_2)$ and
to lowest order in ${d\ln C_l / d\ln \zeta}$ and
$\delta\zeta/\sigma_\zeta$ and obtain 
\beqa\label{cc}
I_l(\nu_1,\nu_2)&\approx& 1 - {1\over 2} \delta \zeta^2
\ln^2(\nu_1/\nu_2) \Biggl \{ {\r \over 1+\r} + {(1+2 \r) \over
  (1+\r)^2}
\left({\delta \zeta \over \sigma_\zeta} \right)^2 \nonumber \\
&+&  \left ( {d\ln C_l \over d\ln \zeta} \delta\zeta \right )^2
\nonumber \\
& & \times \left
     [ {\r \over 4 (1+\r)^2}- {1 +2 \r (1+\r) \over 2 (1+\r)^3}
       \left({\delta \zeta \over \sigma_\zeta} \right)^2  \right ]
     \Biggr \}, 
\eeqa
where $\r$ measures the importance of the Poisson term relative to the
term due to clustering: 
\beqa
\r&=& {C_l^{poisson} \over C_l^{cluster}}, \nonumber \\
C_l^{poisson}&=& \int dS  {dn\over dS } S^2, 
\nonumber \\
C_l^{cluster}&=& C_l^s(\bar \zeta) {\cal I}^2,  \nonumber \\
{\cal I}&=& \int dS {dn\over dS } S.
\nonumber \\
\eeqa
These are the standard formulae for the Poisson and clustering
contributions for a single population of sources
(e.g. \citealt{peebles80}). Equivalently from equation (\ref{bigmess})
and to lowest order in $\delta \zeta/ \sigma_\zeta$ and $d\ln C_l /
d\ln \zeta$ we find 
\beq
C_l(\bar \nu,\bar\nu)= C_l^{poisson}+C_l^s(\bar \zeta) {\cal I}^2 .
\eeq

We can point out a few interesting things about equation
(\ref{cc}). First of all, the departure from unity is proportional to
$\delta \zeta^2 \ln^2(\nu_1/\nu_2)$. As expected, there is a loss of
correlation only to the extent that there are sources with a variety
of spectral indices in the mix.  There are several contributions to
the loss of correlation. The first term on each line (proportional to $\r/
[1+\r]$) is a direct consequence of the Poisson part; they go to
zero as $\r \rightarrow 0$.  They occur because, if the Poisson
contribution is dominant, there is a chance of getting a different
spectral index in different regions of the sky. That is to say, there
are a few sources in any given mode on the sky and so the fluctuations
in the spectral index of actual sources that happen to be in each
region will lead to patterns on the sky that are not identical at
different frequencies. The other terms come from the clustering part
(they go to zero as $\r \rightarrow \infty$). Those terms only appear
if different sources are not perfectly correlated (terms proportional
to $\delta\zeta/ \sigma_\zeta$) or if they cluster differently (terms
proportional to ${d\ln C_l/ d\ln \zeta}$).
  
The main point of equation (\ref{cc}) is that the difference between
the cross-correlation and unity scales as $\delta \zeta^2
\ln^2(\nu_1/\nu_2)$, so it should be quite small. This will imply that
the noise from the smooth foregrounds is unimportant.

\section{The Effect of Contamination}

In this section we will show how the frequency information can be
used to discriminate the 21 cm signal from sources of
contamination. The basic point is that while the contaminants are
presumed to be smooth as a function of frequency, the 21 cm signal
varies very rapidly. A small change in frequency of order a fraction
of a MHz is already enough to sample an effectively different part of
the Universe. In what follows, we will present two derivations that
show that the measurement is only contaminated by the parts of the
foregrounds that are uncorrelated between neighboring frequencies.
For simplicity we will only consider two frequencies and show how the 
combination of information from both can significantly reduce the level 
of contamination. Clearly to make full use of the data set one should 
consider all the frequencies. Here we illustrate the method with just two but 
the generalization is trivial. 

\subsection{Fisher matrix}

To illustrate how the cleaning works let us take a simple model for the data,
\beq
a_{lm}(\nu)= a^{21cm}_{lm}(\nu)+a^{f}_{lm}(\nu)+a^{noise}_{lm}(\nu).
\eeq
The three terms are the 21 cm signal, foreground contamination, and
detector noise.  We will assume that we make $N_l$ measurements at two
separate frequencies, $\nu_1$ and $\nu_2$, so that the data vector is
of the form $\x_i=(a_{lm}(\nu_1),a_{lm}(\nu_2))$ with $i=1,\cdots
N_l$. We will assume that both the 21 cm signal and the detector noise
are uncorrelated between the two frequencies while the foregrounds
have a correlation coefficient $I$, very close to unity. In that case
the correlation matrix of the data is 
\beqa\label{fisher}
\langle \x_i
\x_j^{\dagger}\rangle & = {\bf C}_{ij}= & \delta_{ij} \left [ C^f_l
\left(\begin{array}{cc} 1 & I \sqrt{\beta}\\ \sqrt{\beta} I & \beta \\
\end{array} \right) \right. \nonumber \\ \, & \, & \left. +
C_l^{21cm} \left(\begin{array}{cc}
1 &  0 \\
0 & 1 \\
\end{array} \right) 
+ C_l^N \left(\begin{array}{cc}
1 &  0 \\
0 & 1 \\
\end{array} \right) \right ] ,
\eeqa
where $\beta$ characterizes the frequency dependence of the
foregrounds and $C_l^N$ gives the power spectrum of the noise, which
for simplicity we assumed equal in both channels and uncorrelated
between the different $N_l$ measurements (see \S 7). 

We will assume that the fluctuations are Gaussian and that we wish
to estimate the three parameters of the model,
$\p=(C^f_l,\beta,C_l^{21cm})$, simultaneously. We can calculate the
expected error bars from the Fisher matrix ${\cal F}$, 
\beq
{\cal F}_{i_1,i_2} = {1 \over 2} {\rm Tr}\left [{\bf C}^{-1} {\partial
    {\bf C} \over \partial \p_{i_1}} {\bf C}^{-1} {\partial {\bf C}
    \over \partial \p_{i_2}}\right ]  
\eeq 
where $i_1$, $i_2$ run over the three parameters. The inverse of the
Fisher matrix gives the expected covariance matrix of the recovered
parameters (see Tegmark, Taylor \& Heavens 1997 for a summary of the
Fisher matrix technique). In particular, the error in the recovered 21
cm power spectrum is simply 
\beqa\label{pkerror}
(\Delta C_l^{21 cm})^2& = {\cal F}^{-1}_{3,3}\nonumber &\\ 
&={2 \over N_l}&\left [ \left ( C_l^{21 cm}+ C_l^N + 
{2 \beta \over (1+\beta)}(1-I)C_l^f \right ) ^2 \right. \nonumber \\
\, & \, & \left. +  \left ( {2 \beta
  (1-I)C_l^f  \over (1+\beta)} \right ) ^2\right ], 
\eeqa
where we have assumed  $C_l^f \gg C_l^{21 cm},C_l^N$.

We see that the foreground power spectrum is suppressed by a
factor $1-I$; so, as long as $(1-I) C_l^f <  C_l^{21 cm}$ the noise
introduced by foregrounds can handled with this technique.   

\subsection{Another derivation}

We can obtain the same result as above by considering the following
problem. Assume we measure the $a_{lm}$'s at two different frequencies
and call the results $(x_i,y_i)$ where $i$ runs over the number of
observations. The data is intrinsically of the form 
\beqa
x_i&=&f_i+\epsilon_i, \nonumber \\
y_i&=&\beta^{1/2} f_i + \delta_i \equiv \beta^{1/2} x_i +\mu_i, 
\eeqa
where $f_i$ is the part coming from the correlated foreground
contribution, $\epsilon_i$ has the 21 cm signal plus the noise
contribution at the first frequency and $\delta_i$ has the 21 cm
signal, noise and an additional contribution from the uncorrelated
part of the foregrounds. Just by diagonalizing the covariance matrix
above one can show that this uncorrelated component has variance $2
\beta (1-I) C_l^f$, to lowest order in $(1-I)$.  For convenience we
have introduced $\mu_i=\delta_i-\beta^{1/2} \epsilon_i$. 

The $(x_i,y_i)$ data fall on a straight line with unknown slope
$(\sqrt{\beta})$  that we need to determine. The 21 cm signal is
encoded in the deviations of the data from a perfect line.   We can
determine the slope by writing a simple $\chi^2$ for the best fit
line, 
\beq
\chi^2(b)=\sum_i (y_i-bx_i)^2,
\eeq
where $b$ is the slope that we are trying to determine. We
can easily minimize $\chi^2$ with respect to $b$. The information
about the 21 cm fluctuations is encoded in the value of $\chi^2$ at
this minimum, for which we obtain: 
\beq
\chi^2(b_{min})=\sum_i y_i^2 - {(\sum_i x_i y_i)^2 \over \sum_i x_i^2} .
\eeq
Note that in terms of the underlying model variables,
\beq
\chi^2(b_{min})=\sum_i \mu_i^2 - {(\sum_i x_i \mu_i)^2 \over \sum_i x_i^2} .
\eeq

We can take averages over the uncorrelated component $\mu_i$ (which in
a sense is the only random variable in this approach) to obtain 
\beq
\langle \chi^2(b_{min}) \rangle = (N-1) \sigma_\mu^2.
\eeq
Thus we could use
\beq
\hat S = {1\over N-1} \chi^2(b_{min})
\eeq
as the estimator for the variance, which contains the 21 cm
signal. The mean and variance of this estimator are
\beqa
\langle \hat S \rangle &=& (1+\beta)\left [ C_l^{21 cm}+ C_l^N + 
{2 \beta \over (1+\beta)}(1-I)C_l^f \right ] \nonumber \\
\langle \hat S^2 \rangle-\langle \hat S \rangle^2 &=& {2 (1+\beta)^2
  \over N-1} \left [ C_l^{21 cm}+ C_l^N + 
{2 \beta \over (1+\beta)}(1-I)C_l^f \right ] ^2 .
\eeqa
This is almost the same as we obtained earlier.  The difference can be
traced to the fact that to estimate $C_l^{21 cm}$ from $\hat S$ we
need to subtract the contribution proportional to $(1-I)C_l^f$; this
adds an extra piece to the variance that is second order in $(1-I)C_l^f
$. The important point, however, is that the foreground term appears
only in the form $(1-I)C_l^f $. 

\subsection{The cleaned foreground signal}

\begin{figure}
  \plotone{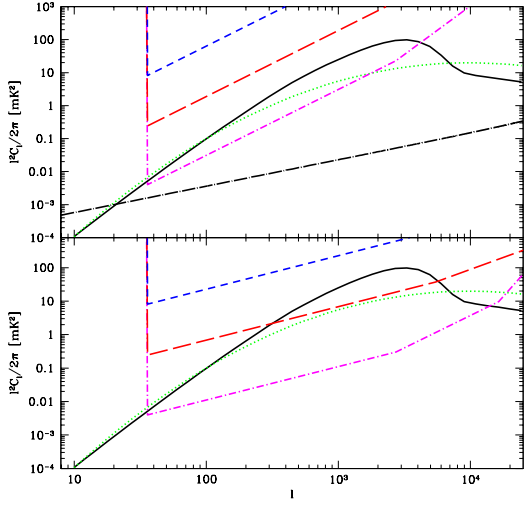}
     \caption{Observability of the estimated 21 cm fluctuation signal.
      The solid black and dotted green curves in each panel are the
      angular power spectrum of the signal at $z=10$ (with
      $\Delta \nu=0.2 \Mhz$) for our toy model and for a fully neutral
      medium.  The long dash-dotted curve in the top panel shows the
      estimated point source foreground signal (see text).  The other
      curves show sensitivity estimates for three proposed
      experiments: PAST (blue short-dashed), LOFAR (red long-dashed),
      and SKA (magenta dot-dashed).  The top panel shows the noise
      power spectrum while the bottom panel shows the error in the
      estimated power spectrum. We assume four weeks of continuous
      observations for each experiment.}
    \label{fig:sensitivity}
\end{figure}

We can now illustrate how the technique we propose reduces the
importance of contamination from unresolved point sources.  
Figure \ref{fig:sensitivity}
again shows the power spectrum at $z=10$, with and without the extra
power from reionization (the other curves in this figure
are described in \S 7).  We also show $(1-I)C_l^f$ for the
``intermediate" point source model of \citet{dimatteo} (dot-dashed
line), assuming that point sources with $S>0.1$ mJy have been
removed. For the correlation coefficient of the maps produced by the
point sources we took $1-I=8\times 10^{-9}$. This corresponds to the
following choices: $\delta \zeta=0.3$, $\ln(\nu_2/\nu_1)=1.3\times
10^{-3}$, $(\delta\zeta/\sigma_\zeta)^2=0.1$, $\r=0$ and $d\ln C_l /
d\ln \zeta=0$.  This choice of $\delta \zeta$ is consistent with the
results of \citet{cohen03}.  Note that we have assumed the Poisson
term to be negligible, as argued by \citet{dimatteo}, and we have
neglected variations in the clustering length with spectral index,
although we do include imperfect correlations between sources with
different $\zeta$.  As the figure shows, the expectation is that once
the frequency information is used, the contamination becomes
significantly smaller than the signal we wish to measure.

\citet{oh03} noted another problem related to simple contamination: if
the beamsize changes with frequency, the number of foreground sources
in a given beam also changes with frequency.  In our language, an
interferometer baseline samples slightly different $l$-modes at
different frequencies.  The most obvious solution is either careful
beamsize control with frequency or fine coverage in Fourier-space
coverage, both of which must be determined during the interferometer
design.  If this is not possible, an alternative method is to note
that this ``leakage'' contamination is also highly correlated in
frequency-space and can be removed with techniques similar to ours
(see also \citealt{gnedin03}).

Note that in our estimate of the foreground removal we have assumed
that the 21 cm signal between the two frequency channels is completely
uncorrelated [i.e. the covariance matrix of the 21 cm signal is
diagonal in equation (\ref{fisher})].  In reality, the off-diagonal
elements of the matrix will contain the cross-correlation of the two
frequency bins, $C_l^{21cm}(\nu_1,\nu_2)$.  The recovered signal is
proportional to $[1-C_l^{21cm}(\nu_1,\nu_2)]$.  This essentially
provides a minimum bandwidth for foreground removal using this
technique: we need the radial separation of the two frequency bins to
exceed the typical correlation length.  In the example we show, we
have chosen $L(\Delta \nu) \ga R$, so the true signal would decrease
by a small amount.  However, the method we have presented uses only
two neighboring frequency channels to remove the contaminants.  In
reality, the full bandwidth of the observation should be used in
estimating the smooth component of the spectrum at each point on the
sky, which will considerably improve the removal algorithm.  This will
allow the single-frequency signal to be isolated without significant
loss from correlations between neighboring frequencies.  We conclude
that point-source foregrounds do not present a significant problem for
these measurements provided that they are smooth in frequency space.

\section{Detectability}

In this section we will examine the prospects for detecting the 21 cm
fluctuations. We will follow the notation of \citet{white99} where the
formalism used to analyze data for CMB interferometers such as the
\emph{Degree Angular Scale Interferometer} (DASI) and the \emph{Cosmic
Background Interferometer} (CBI) was presented.  We note that similar
sensitivity estimates have been made for the \emph{Giant Metrewave
Radio Telescope} by \citet{bharadwaj01} and \citet{bharadwaj03}.

The measured flux in a visibility is
\beq
V(\u)={\partial B_\nu \over \partial T} \int d^2n \ \delta T_b(\n)
A(\n) \ e^{2\pi i \u\cdot\n}, 
\eeq 
where $A(\n)$ is the primary beam and ${\partial B_\nu/\partial T}$
converts temperature to flux. In the Rayleigh-Jeans part of the
spectrum,   ${\partial B_\nu/\partial T}=2 k_B/\lambda^2$. The Fourier
wavenumber $u$ is related to $l$ by $u=l/2\pi$.  We can express
$\delta T_b(\n)$ and  $A(\n)$ in terms of their Fourier components, 
\beqa
\delta T_b(\n)&=&\int d^2u \ \tilde{\delta T}_b(\u) e^{-2\pi i
  \u\cdot\n} \nonumber \\ 
A(\n)&=&\int d^2u \ \tilde{A}(\u) e^{-2\pi i \u\cdot\n}.
\eeqa 
The Fourier components $\tilde{\delta T}_b(\u)$ are essentially the same
as the $a_{lm}$'s introduced earlier except that the spherical harmonic
decomposition has been replaced by a Fourier decomposition, valid only
over small patches of the sky that can be taken to be
flat. The variance of $\tilde{\delta T}_b(\u)$ is given by the power
spectrum,  
\beq
\langle \tilde{\delta T}_b(\u_1) \tilde{\delta T}_b(\u_2)\rangle =
\delta^D(\u_1+\u_2) C_{l=2\pi u_1}. 
\eeq
 In terms of these Fourier variables the variance of the temperature is 
\beq
\langle \delta T_b^2(\n)\rangle = \int d^2 u C_{l=2\pi u}=\int dl {l
  C_l \over 2\pi}, 
\eeq
which can be compared to the exact formula,
\beq
\langle \delta T_b^2(\n)\rangle = \sum_l {(2 l+1)  C_l \over 4\pi}.
\eeq

We now calculate the averaged value of the square of the observed
visibilities in terms of the power spectrum, where the average is over
an ensemble of possible skies, 
\beqa\label{v2}
\langle |V(\u)|^2\rangle&=& \left( {\partial B_\nu \over \partial T}
\right)^2 \int d^2u^\prime \ |\tilde A(\u-\u^\prime )|^2 C_{l=2\pi
  u^\prime} \nonumber \\ 
&\approx& C_{l=2\pi u} \left({\partial B_\nu \over \partial T} \right)^2 \int
d^2u^\prime \ |\tilde A(\u-\u^\prime )|^2. 
\eeqa
If the visibility is observed for a time $t_v$ the averaged noise
squared in each visibility is given by \citep{rohlfs00}
\beq\label{n2}
\langle |N(\u)|^2\rangle =   \left({2 k_B T_{sys} \over
	A_{dish}} \right)^2 \ {1\over \Delta {\nu} t_v}  
\eeq
where  $T_{sys}$ is the system temperature, $\Delta {\nu}$ is the
bandwidth, and $A_{dish}$ is the area of each individual antenna in
the array.

We can compare equations (\ref{v2}) and (\ref{n2}) to define the power
spectrum of the noise, 
\beq
C_l^N= \left ( {2 k_B T_{sys} \over  A_{dish} \partial B_\nu/ \partial T
  } \right )^2 \  {1\over \Delta {\nu} t_v \int d^2u^\prime \
  |\tilde A(\u-\u^\prime )|^2}. 
\eeq
The result of the integral in the denominator depends on the shape of
the primary beam (i.e. the beam of the individual dishes). To get an
approximate answer we can use the fact that $A(\u)$ is different from
zero in an area $d^2u$ and has to integrate to one, so $\int
d^2u^\prime \ |\tilde A(\u-\u^\prime )|^2\sim 1/d^2u$. Moreover, the
size of the primary beam and thus $d^2u$ is directly related to the
area of the dishes. We can approximately use $A_{dish}=\lambda^2
d^2u$.  The power spectrum of the noise then becomes
\beq\label{noise1}
C_l^N= {T_{sys}^2 \over \Delta \nu t_v d^2u}= { T_{sys}^2 \ (2\pi)^2
  \over \Delta \nu t_v d^2l}. 
\eeq
This is equation (17) of White et al. (1999) with the mapping $l=2\pi
u$. Perhaps an easier way to understand this equation is to note that
after a time $t_v$ the noise in the Fourier space pixel corresponding
to the observed visibility is simply $\sigma_n^2=T_{sys}^2/\Delta \nu
t_v$. The noise can be expressed in terms of a power spectrum using
$\sigma_n^2=d^2u C_{l}=d^2l \ C_l/(2 \pi)^2$.  

Interferometric CMB experiments such as DASI and CBI were conducted at
much higher frequencies than those in which we are interested, so the
arrays were small enough that they could be rotated to compensate for
the Earth's rotation.  Consequently, each pair of antennae could
integrate for an arbitrarily long time on a single Fourier component
of the sky. This is not feasible in the present case because the
distances involved are much larger: the arrays must have baselines
on the order of a kilometer.  We therefore need to calculate the
fraction of the total observing time $t_0$ that any given baseline is
being observed. This fraction of time will not be uniform across the
Fourier plane, and the details will depend on the element
configuration.
 
We will make a simple estimate here assuming that the Fourier coverage
is roughly uniform. For a particular maximum separation of the
antennae the interferometer will cover Fourier space up to a maximum
$l_{max}$. Thus, owing to the Earth's rotation, a region of area $\pi
l_{max}^2$ of Fourier space will be covered. At any given instant,
however, only an area $N_{pairs}d^2l$ is being observed. Thus each
visibility will be observed roughly for a time
\beqa\label{noise2}
t_v&\approx& t_o \ {N_{pairs}d^2l \over \pi l_{max}^2} \nonumber \\
&=&  t_o \ {N_{dish}^2 d^2l \over 2 \pi l_{max}^2} ,
\eeqa 
where $N_{pairs}=N_{dish}(N_{dish}-1)/2$ and where we have assumed
$N_{dish} \gg 1$. Combining equations (\ref{noise1}) and
(\ref{noise2}) we obtain
\beq
{l^2 C_l^N \over 2\pi}= {T_{sys}^2 \  (2\pi)^2\over \Delta \nu t_o}\
\left({ l_{max}^2 \over N_{dish} d^2l}\right)^2 \ \left({ l\over
  l_{max}}\right)^2. 
\eeq

We can think of the array as a big telescope with diameter $D$, large
enough to make a measurement of mode $l_{max}$ which covers a total
area $A_{total}$. However, only a fraction of that area is covered
with telescopes, $N_{dish} A_{dish}$. That covering fraction can also
be expressed in terms of $l_{max}$ and $d^2l$, 
\beq
f_{cover}\equiv{N_{dish} A_{dish}\over A_{total}}={ N_{dish} d^2l
  \over l_{max}^2}. 
\eeq
In terms of $f_{cover}$ the noise power spectrum is, 
\beq
{l^2 C_l^N \over 2\pi}= {T_{sys}^2 \ (2\pi)^2 \over \Delta \nu t_o
  f_{cover}^2} \ \left({ l\over l_{max}}\right)^2. 
\eeq

If one is interested in achieving maximum sensitivity at a particular
scale $l$ with a fixed number of elements of a given size, it is best
to pack the elements as close as possible because $ l_{max} \propto D$
but $f_{cover}\propto D^{-2}$. This is achieved when the $l$ of
interest is close to $l_{max}$.  The signal from the ``bubbles" is
located somewhere in the range $l\sim 1000-10000$ so one needs an
array of size  $D=l\lambda/2\pi\sim 300$m - 3 km to observe it.   

For realistic arrays, the distribution of telescopes will not be
uniform, so the coverage in Fourier space will vary with $l$. For
example, the array may have a core at the center where telescopes are
closely packed and a more dilute configuration at larger
separations. Thus the covering fraction for the $l$'s measured by the
core of the array will be much larger than for the higher $l$'s. We can
introduce a function $\tilde f(l)$ that encodes the geometry of the
array, in terms of which
\beq
{l^2 C_l^N \over 2\pi}= {T_{sys}^2 \ (2\pi)^2 \over \Delta \nu t_o
  \tilde f^2(l)} .
\eeq
For our simple case of uniform Fourier coverage,
\beq
\tilde f(l)= f_{cover} {l_{max} \over l }.
\eeq

The system temperature at these frequencies will be dominated by the
sky brightness temperature so the figure of merit to compare different
experiments is simply $\tilde f(l)$. We take $T_{sys}$ to
be roughly $200$ K, so the noise power spectrum is approximately
\beqa
\delta T_n & =& \left [{l^2 C_l^N \over 2\pi}\right ]^{1/2} \nonumber \\
& \sim & 12\ {\rm
  mK} \ \left ({T_{sys}\  \ 0.1\over
  200 K \ \tilde f (l) }\right )\  \left ({0.4 {\rm \ MHz} \ 1 \ {\rm
    month} \over \Delta \nu \ t_o}\right )^{1/2}.  
\eeqa
Thus, if we want an experiment to make a map with good signal to noise
in a matter of weeks it needs to have $\tilde f(l) \sim 0.1$ on the
scales of interest.  Note that this noise estimate does not take
cosmic variance into account. Because each observation samples only a
finite region and must ultimately be compared to a statistical model
of the Universe, at a certain point (when the signal to noise is of
order unity in each Fourier mode), the experimental power becomes
limited by the finite field of view and one is better off increasing
the area of the sky being covered rather than going deeper on the same
spot.  We have quoted the answer for a bandwidth of order $0.4$
MHz. There is little to be gained by making the bandwidth much larger
because by doing so one enters the Limber regime on the angular scales
of interest (see Figure 4 and discussion thereof). Once in the Limber
regime both signal and noise scale as 
$1/\Delta \nu$. The exact crossover into the Limber regime will depend
on the sizes of the bubbles.  On the other hand, if the bandwidth is
much smaller than the typical correlation length of the 21 cm
features, the signal will be difficult to detect as it will be
confounded by the foregrounds. 

If one is interested in a statistical detection of the power spectrum
rather than imaging, the required observing time is significantly
reduced simply because one can make several estimates of the power
spectrum on scales smaller than the total field of view. The error in
the power spectrum estimate is [equation (\ref{pkerror})]
\beqa
{\Delta C_l^{21 cm}\over  C_l^{21 cm}} = \sqrt{2 \over N_l} {C_l^N
  \over C_l^{21 cm}}.
\eeqa
The signal to noise increases by a factor, $\sqrt{N_l/2}\sim \sqrt{
  N_{dish}/ f_{cover}}\sim \sqrt{A_{total}/A_{dish}}$.  In terms of
the $l$-modes sampled, $\sqrt{N_l/2} \sim l/l_{min}$, where $l_{min}$
corresponds to the total angular size of the field of view.

Moreover, we have included only one frequency channel in our estimate
(after foreground subtraction).  If the signal that one is seeking has
a frequency width $\gg 1 \Mhz$, stacking channels adds even more
statistical power: the number of estimates of $C_l^{21cm}$ goes like
the number of channels used in the estimate.  For example, Figure 3
shows that the power spectrum of our toy model changes relatively
little over $\Delta z \sim 1$, corresponding to a frequency width of
$\sim 10 \Mhz$.  Thus, about 25 channels could be stacked without
losing too much redshift information.

Several experiments are now being designed to have the capability to
measure the 21 cm signal. One is a proposed dedicated experiment
called the \emph{Primeval Structure Telescope}\footnote{ See
http://astrophysics.phys.cmu.edu/$\sim$jbp for details on PAST.}
(PAST).  This instrument will have an effective area $N_{dish}
A_{dish} \sim 10^4 {\rm m}^2$ concentrated in a diameter $D\sim 2$ km
($l_{max}\sim 5000$).  For that configuration $\tilde f(l)\sim
N_{dish} A_{dish}/D^2 \ (l_{max}/l) \sim 0.0024 \ l_{max}/l $. Thus
the instrument would require long integrations or averages over many
frequency channels to detect the expected statistical signal.  Note
that some long baselines would also be needed to be able to remove
point source contamination.

Detecting cosmic 21 cm emission is also one of the major science goals
of the \emph{Low Frequency Array}\footnote{ See http://www.lofar.org
for details on LOFAR.} (LOFAR) and the \emph{Square Kilometer
Array}\footnote{ See http://www.skatelescope.org for details on the
SKA.}  (SKA).  LOFAR will have a total effective area of about $2
\times 10^5 $ $ {\rm m^2} $ with approximately 25\% of that area
concentrated in a compact array of $D\sim 2$ km. For this core,
$l_{max}\sim 5000$ and $\tilde f(l) \sim 0.016 \ l_{max}/l$.  The
design of the SKA has not yet been fixed.  Current plans call for
$\sim 20\%$ of the array elements to lie in a core of $D \sim 1$ km
and $\sim 50\%$ to lie within a region of $D \sim 6$ km.  For the
inner region, $l_{max}\sim 2500$ and $\tilde f(l) \sim 0.25 \
l_{max}/l$ and for the outer one $l_{max}\sim 10^{4}$ and $\tilde f(l)
\sim 0.018 \ l_{max}/l$.  Both of these instruments also have the
advantage of very long baselines (hundreds or thousands of kilometers)
that will help with point source removal and control of systematics.

Figure \ref{fig:sensitivity} shows some estimated sensitivity curves
for each of these instruments.  To construct the curves, we assumed
one month of continuous observing on a single field of view.  We
assumed a 100 square degree field of view for each of the experiments.
We have also made some assumptions about the antenna distributions in
these experiments.  For LOFAR, we took ($25\%,\,50\%$) to have
baselines smaller than ($2,\,12$ km).  For SKA, we took
($20\%,\,50\%,\,55\%$) within ($1,\,6,\,12$ km).  We assumed that
$\tilde f(l)$ varies smoothly between these points and uniform Fourier
space coverage within the core.  Note that our sensitivity estimates
are only approximate and depend on the Fourier space coverage of the
array as well as the correlation procedure.

The top panel shows the noise power spectrum; comparison to the power
spectrum of the 21 cm fluctuations gives the signal-to-noise value
for each measured visibility.  This panel therefore shows the
appropriate sensitivity for making a map.  Clearly, creating
images with high signal to noise on arcminute scales will be difficult
and require large collecting areas (on the order of a square
kilometer).  Note that the slope of the sensitivity curves depend on
the antenna configuration; for uniform Fourier coverage, $\delta T_n
\propto l$.  Configurations in which the coverage increases at smaller
separations (i.e. the array's covering fraction increases toward the
center) have steeper slopes.

Fortunately, as noted above, a statistical measurement of the power
spectrum is considerably easier.  In the bottom panel we show the
error on the estimated power spectrum in logarithmic $l$-bins: this is
simply the noise power spectrum multiplied by $N_l^{-1/2} \sim l_{\rm
min}/l$.  The large field of view of SKA allow rather
precise measurements of the power spectrum on scales from $\la 1$
arcmin to $\sim 1\arcdeg$.  Note also that we show the errors in the
individual frequency channels.  If many channels are stacked together,
the errors will decrease by the square root of the number of
channels (ignoring correlations between channels).  In this way, PAST
and especially LOFAR could also make significant detections of the
power spectrum

\section{Conclusions}
\label{sec:conclusion}

The recent successful efforts to map anisotropies in the cosmic
microwave background have determined the global properties of the
Universe to high precision (e.g. Spergel et al. 2003).  When combined
with the highly successful paradigm for the hierarchical growth of
structure (e.g. White \& Rees 1978), these results imply that the
evolution of the Universe on large scales is now relatively
well-understood.  However, on the smaller scales characteristic of
individual objects, many uncertainties remain owing to our ignorance
of the nature of dark matter, the lack of a full physical model for
the origin of primordial density fluctuations, and our poor
understanding of galaxy formation.  Determining the
properties and consequences of the first luminous objects at $z\sim
15-30$ may help to clarify these issues.

The evolution of the ionized part of the IGM at high redshifts
provides a fossil record of the Universe at these times.  In
principle, the physical state of this diffuse gas constrains when and
where the first luminous objects formed as well as the nature of the
sources responsible for reionization.  The relatively large electron
scattering optical depth obtained from the WMAP observations is
indicative of a complex reionization history but, by itself, does
not provide unambiguous answers to the remaining questions.

In this paper, we have argued that fluctuations in 21 cm emission from
the IGM can be used to measure the three-dimensional distribution of
neutral hydrogen at high redshifts.  Measurements of the angular power
spectrum at different frequencies can be used to mitigate
contamination by foreground sources that would otherwise overwhelm the
21 cm signal.  Our approach is similar to that employed in analyses of
CMB anisotropies, but is more general because of the
frequency-dependent nature of redshifted 21 cm emission, making it
possible to use 21 cm fluctuations study the {\it evolution} of
reionization.
We note here that the most basic kind of experiment is to seek an
``all-sky'' signature corresponding to a global phase transition in
the neutral hydrogen (which could be reionization, reheating, or the
onset of Ly$\alpha$ coupling; \citealt{shaver}).  While valuable
(especially because they have much less stringent sensitivity
requirements), these measurements only provide the most basic
information.  Moreover, they are subject to severe foreground
contamination \citep{gnedin03} because the large scale signal varies
relatively smoothly with frequency.  Thus frequency differencing will
not be as effective as with small-scale fluctuations.  The small-scale
fluctuations we have described will therefore ultimately be a better
route to pursue.

Using a simple conceptual model for the morphological evolution of
ionized regions, we have demonstrated how reionization imprints
characteristic features into the angular power spectrum of 21 cm
fluctuations.  Our formulation is general, but for illustrative
purposes we have adopted simplifying assumptions to emphasize salient
features of our methodology.  For example, in our toy model of
reionization, we have ignored redshift space distortions in the 21 cm
signal and neglected correlations between the density and neutral
fraction fields.  In future work, we will investigate the impact of
these effects explicitly using semi-analytic methods and numerical
simulations to show how the formalism can be adapted to account for
these complications.

In principle, measurements of the type we propose can be used to
distinguish between various evolutionary histories.  Models with
multiple epochs of reionization (e.g. Cen 2003; Wyithe \& Loeb 2003)
lead to a behavior in which the volume fraction of ionized gas in
the universe shows a complex dependence on redshift (e.g. Figures
8 \& 9 of Sokasian et al. 2003b), unlike single episodes of
reionization where the trend is simpler (e.g. Figure 5 of Sokasian
et al. 2003a).  These differences will be imprinted onto the
frequency dependence of 21 cm fluctuations of the IGM and can
be discerned using the approach described here.  
In fact, the simple model described in \S 4 is only part of the story
available to us through 21 cm observations.  For example, the
fluctuations constrain the thermal history of the IGM as well as the
ionization history (see \S 2).  If the earliest ionizing sources have
hard spectra (such as quasars), we would expect the IGM to be heated
rapidly, while if cool, low-mass stars are responsible for
reionization, the heating would occur much closer to overlap.  Another
way to look at this is that the reionization history determines how
much information is available to us through 21 cm fluctuations.  

For instance, in a scenario with rapid heating, we will have a long
epoch where density variations dominate the signal.
The power spectrum of the 21 cm fluctuations is a direct tracer of the
matter power spectrum. 
Thus, detailed measurements of this signal could
provide invaluable constraints on the primordial spectrum on small
scales which could tightly constrain the physics of the early
Universe. In particular, because one gets many independent maps by
varying the frequency, the constraints on the power spectrum obtained
in this way could be significantly better than those coming from the CMB
primary anisotropies.
Also, 21 cm fluctuations allow one to probe the power spectrum at $z
\sim 10$--$20$, an epoch inaccessible to both the CMB and galaxy
surveys and a relatively large lever with which to study the growth of
fluctuations.  On the other hand, the 21 cm signal depends on complex
physical processes (see \S 2) and interpreting the results will
require careful modeling.

Differences in the evolutionary state of the ionized IGM also contain
information about the matter power spectrum on small scales.  For
example, in cosmological models with reduced small-scale power, the
halos hosting star-forming regions form late, delaying the ionizing
effects of the first luminous objects (e.g. Somerville et al. 2003;
Yoshida et al. 2003b,c).  Thus, universes with a large component of
warm dark matter or those in which the matter power spectrum has a
running spectral index should exhibit lower amplitude fluctuations in
21 cm emission from higher redshifts than in conventional $\Lambda$CDM
models.

A detailed study of 21 cm fluctuations can also constrain the nature
of the sources responsible for reionization.  In some scenarios, it is
conjectured that reionization occurs ``outside-in,'' affecting voids
first and high density regions later (e.g. Miralda-Escud\'e et
al. 2000).  This progression would be expected if the sources are
rare, but bright.  Alternatively, reionization could proceed in the
opposite sense, ``inside-out,'' particularly if the sources are
numerous, but faint (e.g. Gnedin 2000; Sokasian et al. 2003a).  The
former possibility would apply if quasars were the primary sources of
ionizing radiation, while stars in low-mass galaxies would be more
relevant to the ``inside-out'' scenario.  The morphological appearance
of the ionized gas is different in these two cases, and these
differences would be reflected in the angular power spectrum of 21 cm
fluctuations and how this quantity varies with frequency.
In particular, the size of the \ion{H}{2} regions at a given $x_H$
constrains the number density of ionizing sources.  Also, if voids are
ionized first the density and $x_H$ fields will be correlated (i.e.,
the neutral fraction falls first in regions that are already
underdense), while the opposite is true in an ``inside-out'' scenario.
Thus the amplitude of the brightness temperature fluctuations contains
information about the process of reionization.

As we have argued in \S 7, the technological requirements for
detecting 21 cm fluctuations from cosmic gas at high redshifts, while
demanding, are within reach.  In the near future, it is likely that
measurements of the power spectrum of 21 cm fluctuations will reveal
the physical state of the Universe at an epoch that is currently
inaccessible to other observational probes.

\acknowledgments 

We would like to thank Frank Briggs, Marc Kamionkowski, Antony Lewis,
Avi Loeb, Ue-Li Pen, Jeff Peterson, Scott Schnee, and Peter Sollins
for useful discussions. This work was supported in part by NSF grants
ACI AST 99-00877, AST 00-71019, AST 0098606, and PHY 0116590 and NASA
ATP grants NAG5-12140 and NAG5-13292 and by the David and Lucille
Packard Foundation Fellowship for Science and Engineering.


\begin{thebibliography}
\expandafter\ifx\csname natexlab\endcsname\relax\def\natexlab#1{#1}\fi

\bibitem[{{Allison} \& {Dalgarno}(1969)}]{allison}
{Allison}, A.~C., \& {Dalgarno}, A. 1969, \apj, 158, 423

\bibitem[{{Barkana} \& {Loeb}(2001)}]{barkana01}
{Barkana}, R., \& {Loeb}, A. 2001, \physrep, 349, 125

\bibitem[{{Becker} {et~al.}(2001)}]{becker}
{Becker}, R.~H., {et~al.} 2001, \aj, 122, 2850

\bibitem[Bharadwaj \& Pandey(2003)]{bharadwaj03} 
Bharadwaj, S.~\& Pandey, S.~K.\ 2003, Journal of Astrophysics and
Astronomy, 24, 23

\bibitem[Bharadwaj \& Sethi(2001)]{bharadwaj01} 
Bharadwaj, S.~\& Sethi, S.~K.\ 2001, Journal of Astrophysics and
Astronomy, 22, 293

\bibitem[{{Carilli} {et~al.}(2002){Carilli}, {Gnedin}, \& {Owen}}]{carilli}
{Carilli}, C.~L., {Gnedin}, N.~Y., \& {Owen}, F. 2002, \apj, 577, 22

\bibitem[{{Cen}(2003)}]{cen03}
{Cen}, R. 2003, \apjl, 591, L5

\bibitem[{{Chen} \& {Miralda-Escud{\' e}}(2003)}]{chen03}
{Chen}, X., \& {Miralda-Escud{\' e}}, J. 2003, \apj, submitted,
  [astro-ph/0303395]

\bibitem[{{Ciardi} \& {Madau}(2003)}]{ciardi03}
{Ciardi}, B., \& {Madau}, P. 2003, \apj, 596, 1

\bibitem[{{Cohen} {et~al.}(2003)}]{cohen03} {Cohen}, A.~S.,
{Rottgering}, H.~J.~A., {Jarvis}, M.~J., {Kassim}, N.~E., \& {Lazio},
T.~J.~W. 2003, \apj, in press [astro-ph/0310521]

\bibitem[{{Cooray}(2003)}]{cooray03}
{Cooray}, A. 2003, astro-ph/0309301

\bibitem[{{Couchman} \& {Rees}(1986)}]{couchman86}
{Couchman}, H.~M.~P., \& {Rees}, M.~J. 1986, \mnras, 221, 53

\bibitem[{{Di Matteo} {et~al.}(2002){Di Matteo}, {Perna}, {Abel}, \&
  {Rees}}]{dimatteo}
{Di Matteo}, T., {Perna}, R., {Abel}, T., \& {Rees}, M.~J. 2002, \apj, 564, 576


\bibitem[{{Fan} {et~al.}(2002)}]{fan}
{Fan}, X., {et~al.} 2002, \aj, 123, 1247

\bibitem[{{Field}(1958)}]{field58}
{Field}, G.~B. 1958, Proc. IRE, 46, 240

\bibitem[{{Field}(1959{\natexlab{a}})}]{field59a}
---. 1959{\natexlab{a}}, \apj, 129, 525

\bibitem[{{Field}(1959{\natexlab{b}})}]{field59b}
---. 1959{\natexlab{b}}, \apj, 129, 551

\bibitem[{{Furlanetto} \& {Loeb}(2002)}]{furl-21cm}
{Furlanetto}, S.~R., \& {Loeb}, A. 2002, \apj, 579, 1

\bibitem[{{Furlanetto} {et~al.}(2003){Furlanetto}, {Sokasian}, \&
  {Hernquist}}]{furl-21cmsim}
{Furlanetto}, S.~R., {Sokasian}, A., \& {Hernquist}, L. 2003, \mnras, in press
  [astro-ph/0305065]

\bibitem[{{Gnedin}(2000)}]{gned00} {Gnedin}, N.Y. 2000, \apj, 535, 530

\bibitem[{{Gnedin} \& {Shaver}(2003)}]{gnedin03} {Gnedin}, N.~Y. \&
  Shaver, P.~A. 2003, \apj, submitted (astro-ph/0312005)

\bibitem[{{Gruzinov} \& {Hu}(1998)}]{gruzinov98}
{Gruzinov}, A., \& {Hu}, W. 1998, \apj, 508, 435

\bibitem[{{Gunn} \& {Peterson}(1965)}]{gp}
{Gunn}, J.~E., \& {Peterson}, B.~A. 1965, \apj, 142, 1633

\bibitem[{{Haiman} \& {Holder}(2003)}]{haiman03}
{Haiman}, Z., \& {Holder}, G.~P. 2003, \apj, 595, 1

\bibitem[{{Holder} {et~al.}(2003){Holder}, {Haiman}, {Kaplinghat}, \&
  {Knox}}]{holder03}
{Holder}, G.~P., {Haiman}, Z., {Kaplinghat}, M., \& {Knox}, L. 2003, \apj, 595,
  13

\bibitem[{{Hu}(1999)}]{hu99}
{Hu}, W. 1999, \apjl, 522, L21

\bibitem[{{Hu} \& {Dodelson}(2002)}]{hu02}
{Hu}, W., \& {Dodelson}, S. 2002, \araa, 40, 171

\bibitem[{{Hu} \& {Holder}(2003)}]{hu03}
{Hu}, W., \& {Holder}, G.~P. 2003, \prd, 68, 23001

\bibitem[{{Hui} \& {Haiman}(2003)}]{hui03}
{Hui}, L., \& {Haiman}, Z. 2003, \apj, 596, 9

\bibitem[{{Iliev} {et~al.}(2002){Iliev}, {Shapiro}, {Ferrara}, \&
{Martel}}]{iliev}
{Iliev}, I.T., {Shapiro}, P.R., {Ferrara}, A. \&
{Martel}, H. 2002, \apj, 572, L123

\bibitem[Kaiser(1992)]{kaiser92} Kaiser, N.\ 1992, \apj, 388, 
272 

\bibitem[{{Kaplinghat} {et~al.}(2003){Kaplinghat}, {Chu}, {Haiman}, {Holder},
  {Knox}, \& {Skordis}}]{kaplinghat03}
{Kaplinghat}, M., {Chu}, M., {Haiman}, Z., {Holder}, G.~P., {Knox}, L., \&
  {Skordis}, C. 2003, \apj, 583, 24

\bibitem[{{Knox} {et~al.}(1998){Knox}, {Scoccimarro}, \& {Dodelson}}]{knox98}
{Knox}, L., {Scoccimarro}, R., \& {Dodelson}, S. 1998, Physical Review Letters,
  81, 2004

\bibitem[{{Kogut} {et~al.}(2003)}]{kogut03}
{Kogut}, A., {et~al.} 2003, \apjs, 148, 161

\bibitem[{{Kumar} {et~al.}(1995){Kumar}, {Padmanabhan}, \&
  {Subramanian}}]{kumar}
{Kumar}, A., {Padmanabhan}, T., \& {Subramanian}, K. 1995, \mnras, 272, 544

\bibitem[{{Limber}(1953){Limber}}]{limber} {Limber}, D.N. 1953, \apj, 117, 134

\bibitem[{{Mackey} {et~al.}(2003){Mackey}, {Bromm}, \& {Hernquist}}]{mackey03}
{Mackey}, J., {Bromm}, V., \& {Hernquist}, L. 2003, \apj, 586, 1

\bibitem[{{Madau} {et~al.}(1997){Madau}, {Meiksin}, \& {Rees}}]{mmr}
{Madau}, P., {Meiksin}, A., \& {Rees}, M.~J. 1997, \apj, 475, 429

\bibitem[{{Miralda-Escud{\' e} {et~al.}}(2000)}]{metal00}
{Miralda-Escud{\' e}}, J., {Haehnelt}, M., \& {Rees}, M. 2000, \apj, 530, 1

\bibitem[{{Oh} \& {Mack}(2003)}]{oh03}
{Oh}, S.~P., \& {Mack}, K.~J. 2003, \mnras, submitted, [astro-ph/0302099]

\bibitem[{{Peebles}(1980)}]{peebles80} {Peebles}, P.~J.~E. 1980, {The
Large-Scale Structure of the Universe} (Princeton: Princeton
University Press)

\bibitem[{{Pen}(2003)}]{pen03}
{Pen}, U. 2003, astro-ph/0305387

\bibitem[Rohlfs \& Wilson(2000)]{rohlfs00} Rohlfs, K.~\& Wilson,
T.~L.\ 2000, Tools of Radio Astronomy (New York: Springer)

\bibitem[{Santos} {et~al.}(2003)]{santos03} Santos, M.~G., Cooray, A.,
Haiman, Z., Knox, L., \& Ma, C.-P. 2003, \apj, in press
[astro-ph/0305471]

\bibitem[{{Scott} \& {Rees}(1990)}]{scott}
{Scott}, D., \& {Rees}, M.~J. 1990, \mnras, 247, 510

\bibitem[{{Shaver} {et~al.}(1999){Shaver}, {Windhorst}, {Madau}, \& {de
  Bruyn}}]{shaver}
{Shaver}, P.~A., {Windhorst}, R.~A., {Madau}, P., \& {de Bruyn}, A.~G. 1999,
  \aap, 345, 380

\bibitem[{{Sokasian} {et~al.}(2001){Sokasian}, {Abel}, \&
  {Hernquist}}]{sok01}
{Sokasian}, A., {Abel}, T., \& {Hernquist}, L. 2001, NewA, 6, 359

\bibitem[{{Sokasian} {et~al.}(2002){Sokasian}, {Abel}, \&
  {Hernquist}}]{sokasian02}
{Sokasian}, A., {Abel}, T., \& {Hernquist}, L. 2002, \mnras, 332, 601

\bibitem[{{Sokasian} {et~al.}(2003a){Sokasian}, {Abel}, {Hernquist}, \&
  {Springel}}]{sokasian03a}
{Sokasian}, A., {Abel}, T., {Hernquist}, L., \& {Springel}, V. 2003a, \mnras,
  344, 607

\bibitem[{{Sokasian} {et~al.}(2003b){Sokasian}, {Abel}, {Hernquist}, \&
  {Springel}}]{sokasian03b}
{Sokasian}, A., {Abel}, T., {Hernquist}, L., \& {Springel}, V. 2003b, \mnras,
submitted [astro-ph/0307451]

\bibitem[{{Somerville} {et~al.}(2003)}]{somer03}
{Somerville}, R.S., {Bullock}, J.S., \& {Livio}, M. 2003, \apj, 593, 616

\bibitem[{{Spergel} {et~al.}(2003)}]{spergel03}
{Spergel}, D.~N., {et~al.} 2003, \apjs, 148, 175

\bibitem[{{Springel} \& {Hernquist}(2003a){Springel} \& {Hernquist}}]{sh03a}
{Springel}, V. \& {Hernquist}, L. 2003a, \mnras, 339, 289

\bibitem[{{Springel} \& {Hernquist}(2003b){Springel} \& {Hernquist}}]{sh03b}
{Springel}, V. \& {Hernquist}, L. 2003b, \mnras, 339, 312

\bibitem[Tegmark, Taylor, \& Heavens(1997)]{1997ApJ...480...22T} Tegmark, 
M., Taylor, A.~N., \& Heavens, A.~F.\ 1997, \apj, 480, 22 

\bibitem[{{Theuns} {et~al.}(2002)}]{theuns02-reion}
{Theuns}, T., {et~al.} 2002, \apjl, 567, L103

\bibitem[{{Tozzi} {et~al.}(2000){Tozzi}, {Madau}, {Meiksin}, \& {Rees}}]{tozzi}
{Tozzi}, P., {Madau}, P., {Meiksin}, A., \& {Rees}, M.~J. 2000, \apj, 528, 597

\bibitem[{{Venkatesan} {et~al.}(2001){Venkatesan}, {Giroux}, \&
  {Shull}}]{venkatesan01}
{Venkatesan}, A., {Giroux}, M.~L., \& {Shull}, J.~M. 2001, \apj, 563, 1

\bibitem[{{White}, \& {Rees}(1978)}]{white78} {White}, S.D.M., \&  {Rees}, M.~J.\ 1978, 
\mnras, 183, 341 

\bibitem[White, Carlstrom, Dragovan, \& 
Holzapfel(1999)]{white99} White, M., Carlstrom, J.~E., 
Dragovan, M., \& Holzapfel, W.~L.\ 1999, \apj, 514, 12 

\bibitem[{{Wouthuysen}(1952)}]{wout}
{Wouthuysen}, S.~A. 1952, \aj, 57, 31

\bibitem[{{Wyithe} \& {Loeb}(2003)}]{wyithe03}
{Wyithe}, J.~S.~B., \& {Loeb}, A. 2003, \apjl, 588, L69

\bibitem[{{Yoshida} {et~al.}(2003a){Yoshida}, {Abel}, {Hernquist}, \&
  {Sugiyama}}]{yoshida03-semian}
{Yoshida}, N., {Abel}, T., {Hernquist}, L., \& {Sugiyama}, N. 2003a, \apj, 592,
  645

\bibitem[{{Yoshida} {et~al.}(2003b){Yoshida}, {Sokasian}, {Hernquist}, \&
  {Springel}}]{yoshida03b}
{Yoshida}, N., {Sokasian}, A., {Hernquist}, L., \& {Springel}, V. 2003b, \apj, 591, L1

\bibitem[{{Yoshida} {et~al.}(2003c){Yoshida}, {Sokasian}, {Hernquist}, \&
  {Springel}}]{yoshida03c}
{Yoshida}, N., {Sokasian}, A., {Hernquist}, L., \& {Springel}, V. 2003c, \apj, in press [astro-ph/0305517]

\bibitem[{{Yoshida} {et~al.}(2003d){Yoshida}, {Bromm}, \& {Hernquist}}]{ybh}
{Yoshida}, N., {Bromm}, V., \& {Hernquist}, L. 2003d, \apj, submitted
[astro-ph/0310443]

\bibitem[Zaldarriaga(1997)]{zal97}
Zaldarriaga, M.\ 1997,  \prd, 55, 1822 

\end{thebibliography}
\end{document}